\documentclass{aa}  
\let\linenumbers\relax
\let\endlinenumbers\relax

\usepackage{graphicx}
\usepackage{txfonts}
\usepackage{lipsum}
\usepackage{subcaption}         
\usepackage{lscape}             
\usepackage{placeins}           

\usepackage{lineno}
\nolinenumbers  
\usepackage{txfonts}
\usepackage{lineno}  
\nolinenumbers       

\usepackage[colorlinks=true,linkcolor=blue,citecolor=blue,urlcolor=blue]{hyperref}%
\usepackage[dvipsnames]{xcolor}

\newcommand{\hi}{\ion{H}{i}}
\newcommand{\hii}{\ion{H}{ii}}

\newcommand{\lya}{Ly$\alpha$\xspace}
\newcommand{\oiii}{O\,{\sc iii}}
\newcommand{\oii}{\ion{O}{ii}}

\newcommand{\jwst}{\emph{JWST}\xspace}
\newcommand{\src}{CAPERS-UDS-32520\xspace}

\begin{document}

   \title{A hot, nebular-dominated galaxy interacting with a candidate pristine PopIII system uncovered by \jwst}



   \author{Henriette~Reumert\inst{1,2}
   \and
   Kasper~E.~Heintz\inst{3,1,2}
   \and
   Clara~L.~Pollock\inst{1,2}
   \and
   Alex~J.~Cameron\inst{1,2}
   \and
   Gabriel~B.~Brammer\inst{1,2}
   \and 
   Albert~Sneppen\inst{1,2}
   \and
   Joris~Witstok\inst{1,2}
   \and 
   Chamilla~Terp\inst{1,2}
   \and
   Darach~Watson\inst{1,2}} 
        
   \institute{Cosmic Dawn Center (DAWN), Denmark; 
    \email{henriette.reumert@nbi.ku.dk}
    \and Niels Bohr Institute, University of Copenhagen, Jagtvej 128, 2200 Copenhagen N, Denmark
    \and DTU Space, Technical University of Denmark, Elektrovej 327, DK2800 Kgs. Lyngby, Denmark
    }

   \date{Received \today}

 
    \abstract{
    The discovery of galaxies with extremely strong nebular continuum emission at high redshifts provide novel, unique insights into the conditions under which the first super-massive stars formed. Here we identify a galaxy at redshift $z=5.124$ observed by the \jwst\ CAPERS survey that exhibits a prominent turnover in the rest-frame UV continuum and a pronounced Balmer `jump'. We model the entire \jwst/NIRSpec Prism spectrum from rest-frame UV to optical wavelength, finding that a dominant ($>95\%$) nebular continuum emission can accurately reproduce the spectral shape across all wavelengths. We tested an alternative model with strong damped \lya\ absorption (DLA), but found that it is not able to match the shape of the turnover without invoking a large freedom in the redshift of the absorber. The nebular continuum emission model reveals a hot ($T = (5.3\pm 0.2)\times 10^{4}$\,K) and dense ($n_e = (5.4\pm 0.8)\times 10^{3}\,{\rm cm^{-3}}$) nebular region powering the origin of the spectral shape. We also note the presence of a `blue' candidate companion source potentially at the same redshift, offset by 3\,kpc to the main galaxy. Intriguingly, the spectrum of this source show several hints of hydrogen and helium lines, but no metal lines are detected. We theorize that this companion galaxy might be comprised mainly of Population III (PopIII) stellar remnants and potentially powers the nebular continuum emission seen in the main galaxy. These results have important implications for the presence of a potential dominant population of super-massive and PopIII stars and their consequent excess UV brightness for a significant fraction of galaxies at cosmic dawn.
    }

   \keywords{galaxies: ISM – galaxies: starburst – galaxies: star formation - galaxies: high redshift}

    \titlerunning{An extreme nebular-dominated galaxy at $z=5.124$}
    \authorrunning{Reumert et al.}

   \maketitle

\section{Introduction}

The advent of the {\em James Webb Space Telescope} \citep[\jwst,][]{Gardner06} has provided unprecedented constraints on the rest-frame UV and optical properties of galaxies within the first billion years of cosmic time at redshifts $z\gtrsim 5$. 
One of the most significant challenges raised by \jwst observations is the reported overabundance of UV-bright galaxies at $z\gtrsim 10$ \citep[e.g.,][]{Harikane2024, Whitler2025}.
On the one hand, this could suggest a transition in the mode of star formation to environments with enhanced star formation efficiency \citep[][]{Dekel2023} and stochastically varying (`bursty') star-formation histories \citep[][]{Ren2019, Mason2023, Sun2023, Gelli2024, Munoz2026}. On the other hand, it could signal variations in the mass-to-light ratio of stellar populations due to a stellar initial mass function (IMF) with a greater fraction of massive stars (`top-heavy' e.g., \citealt{2023Steinhardt,Yung2024}), or a boosting of the UV luminosity due to a stronger nebular continuum contribution than previously expected \citep[][]{Cameron2024_NDG, Katz25}.

The latter scenario occurs when ionizing photons are absorbed by the gas and then reprocessed into continuum emission through three different processes: free-free, free-bound, and two-photon ($2\gamma$, see e.g. fig. 1 of \citealt{Katz25} for a schematic). 
The free-bound nebular continuum, whose impact on galaxy SEDs is sometimes observed in the form of a `Balmer jump' discontinuity at the wavelength of the Balmer limit,  is often observed in young, low-mass, metal-poor galaxies at $z\sim0$ \citep[][]{Peimbert1969,Guseva2006,Guseva2007}.
Numerical simulations have predicted Balmer jumps to be increasingly common at high redshift \citep[][]{Katz2023_SPHINX_DR,Wilkins2024}, and a growing number of spectroscopic measurements of Balmer jumps have been made with \emph{JWST}/NIRSpec \citep[][]{RobertsBorsani2024, Cameron2024_NDG, Katz25}. Meanwhile, insights from medium-band imaging suggest that as many as $\sim$30~\% of galaxies may exhibit Balmer jumps by $z\sim6$ \citep[][]{Trussler2025}.

\begin{figure*}[!t]
    \centering
    \includegraphics[width=18cm]{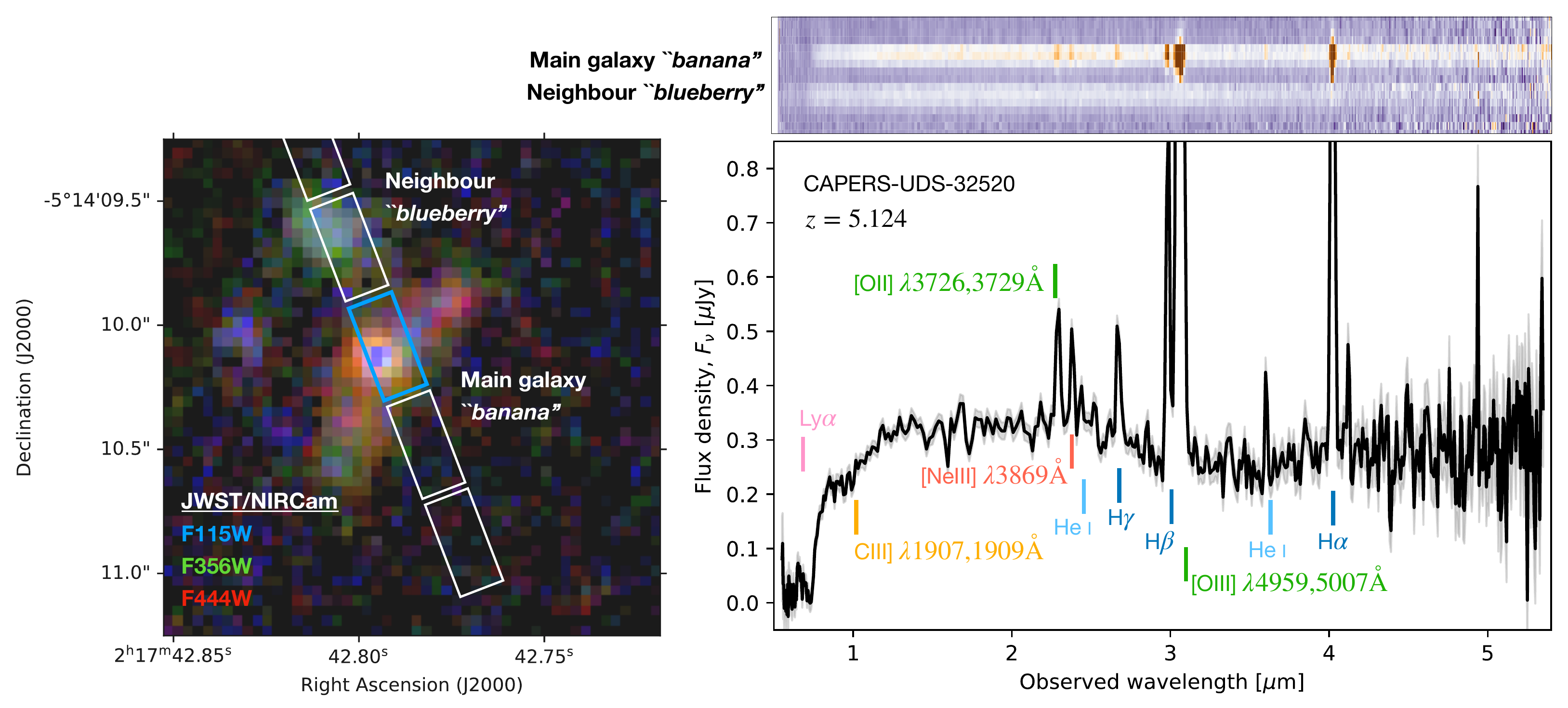}
    \caption{({\it Left:}) JWST/NIRcam false-color red-green-blue (RGB) thumbnail image, showing a $2''\times 2''$ cutouts centred on the main target (the {\it ``banana''}) from the {\tt Primer} survey \citep{Donnan24}.   
    ({\it Right:}) The top shows the 2D spectrum extracted from the shutter, where the bottom shows the 1D spectrum extracted using a global-average background subtraction to remove potential contamination from the nearby neighbor (the {\it ``blueberry''}, see text for details). The most prominent emission-line features identified at the spectroscopic redshift $z=5.124$ are marked. }
    \label{fig:spec}
\end{figure*}

In particular, \citet{Cameron2024_NDG} and \citet{Katz25} reported several examples of so-called `nebular dominated galaxies' at $z>4.5$, for which both the free-bound Balmer jump in the optical and the continuum modeling suggests strong two-photon continuum. Consequently, the nebular continuum contributes $\gtrsim60-80$~\% of the rest-frame UV luminosity, with the rest originating from stellar sources.
Such high nebular contribution cannot be achieved with standard stellar population models, with those authors suggesting a significant contribution from very hot stars ($T_{\rm eff}\gtrsim80,000$~K), perhaps indicating exotic stellar populations and/or a very top-heavy IMF.
Such a strong nebular continuum is inferred based on the combination of a clear Balmer jump, a strong UV rollover, peaking at $1420$Å due to $2$-photon emission, and nebular emission lines with high equivalent widths \citep[EWs, though see e.g.][for alternative interpretations]{Terp2024,Li2024,Tacchella2025}.
If these systems truly are powered by such extreme stellar populations, this clearly has significant implications for how the UVLF can be interpreted.

The identification of nebular-dominated galaxies and the robustness with which they can be confirmed is limited by the emergence of a large population of galaxies with strong damped Lyman-$\alpha$ absorption features ($N_{\rm HI}\gtrsim 10^{21}\,$cm$^{-2}$) at similar redshifts \citep[e.g.,][]{Heintz24,Heintz25,Umeda24,Hainline24,Terp2024,Witstok25}, mimicking the predicted UV `rollovers' of nebular-dominated galaxies at a given temperature. \citet{Katz25} carried out a more systematic search for high-redshift galaxies with pronounced Balmer `jumps' and provided theoretical predictions for their rest-frame UV and optical characteristics easing their identification.

Here we report the identification of a remarkable galaxy at $z=5.124$ observed as part of the \jwst/NIRSpec Prism CANDELS-Area Prism Epoch of Reionization Survey \citep[CAPERS;][]{Dickinson24}, henceforth denoted as `\src'. This galaxy is characterized by a very pronounced rest-frame UV `rollover' and flux excess relative to the optical continuum, exemplary of a strong Balmer jump. Notably, this galaxy does not show a Ly$\alpha$ emission feature expected for nebular-continuum-dominated galaxies and observed previously at high-redshifts \citep[][]{Cameron2024_NDG,Katz25}. This motivated our more detailed analysis and investigation into the underlying physical mechanisms driving the observed continuum and emission-line flux of \src. 

We have structured the manuscript as follows: In Sect.~\ref{sec:data}, we detail the \jwst\ imaging and spectroscopic observations of \src and in Sect.~\ref{sec:res} we present the analysis and results of the spectroscopic modeling. In Sect.~\ref{sec:uvroll}, we examine the various likely causes for the remarkable continuum shape, and in Sect.~\ref{sec:popiii} we discuss the relevance of the potential pristine, PopIII-dominated companion galaxy detected in the image. In Sect.~\ref{sec:conc} we discuss and conclude on our work. Throughout the paper, we assume a standard $\Lambda$CDM concordance cosmology, with parameters adopted from \citet{Planck18} and applied via the {\tt AstroPy} Python software \citep{Astropy} for cosmological inferences.  


\section{Observations and data processing} \label{sec:data} 

The spectroscopic observations of \src were obtained using the NIRSpec Prism configuration of \jwst with approximate resolving power $\mathcal{R} \sim 100$ and wavelength coverage $\lambda \sim 0.6-5.5\mu$m \citep[][]{Jakobsen22} as part of the CAPERS \jwst\ Cycle 3 program \citep[Prog. ID: 6368,][]{Dickinson24}. The source is located at right ascension and declination, $\alpha,\delta ({\rm J}\,2000) = 02^{\rm h}17^{\rm m}42^{\rm s}79, -05^{\circ}14^{'}10\farcs 2$ and has an absolute UV magnitude of $M_{\rm UV} = -20.95\pm 0.05$\,mag. The main galaxy properties are summarized in Table \ref{tab:prop}. For this work, we utilize the reduced and processed spectra of \src from the DAWN JWST Archive (DJA)\footnote{\url{https://dawn-cph.github.io/dja/}}. The details of these reductions are described in separate works \citep{Heintz25,DeGraaff24}. Briefly, the raw spectroscopic data are retrieved from MAST, before they are processed with {\tt MSAExp} \citep[][]{Brammer_msaexp}, creating homogeneous, science-ready data products. Here we use the DJA-{\tt v4} spectra, which extends the nominal wavelength coverage of NIRSpec Prism up to $5.5\mu$m and includes an improved bar shadow correction \citep[][]{Pollock25}. The absolute flux calibration and color corrections of this version of the spectra are typically within $\lesssim 10\%$ of the photometrically-determined values at all wavelengths. 

\begin{table}
    \begin{center}
     \renewcommand{\arraystretch}{1} 
            \caption{Overview of the properties of \src.}
        \begin{tabular}{ l r} 
         \hline \hline    
            R.A. (J2000) & $02^{\rm h}17^{\rm m}42^{\rm s}79$ \\
            Decl. (J2000) & $-05^{\circ}14^{'}10\farcs 2$ \\
            $z_{\rm spec}$ & $5.1240\pm 0.0002$ \\
            $M_{\rm UV}$ (mag) & $-20.95\pm 0.05$ \\
            $\beta_{\rm UV}$  & $-1.68\pm 0.05$ \\
            \vspace{0.3cm}
            $D_{\rm Ly\alpha}$ & $61.3\pm 6.2\,\AA$ \\
            Nebular model & \\
            \hline   
            $T$ (K) & $(5.3 \pm 0.2) \times 10^4$ \\
            $n_e$ & $(5.4 \pm 0.8) \times 10^3$ \\
            \hline \hline  
        \end{tabular}  
        \label{tab:prop}
        \end{center}
    
    
\end{table}

The final extracted 1D and 2D spectra are shown in Fig.~\ref{fig:spec}. For the shown versions, we performed a global sky background subtraction (not included in the online DJA repository) due to the fainter blue companion galaxy seen in the left panel of Fig.~\ref{fig:spec}. 

The \jwst\ imaging for \src\ were obtained from the public available \jwst\ Public Release IMaging for Extragalactic Research ({\tt Primer}) survey \citep[e.g.,][GO-1837]{Dunlop21,Donnan24}, delivering deep \jwst\ NIRCam+MIRI imaging of the COSMOS and UDS legacy fields. For this work, we use the reduced images and photometric catalogs publicly available on DJA \citep[see e.g.,][]{Valentino23}. The main source (the {\it `banana'}) is shown in Fig.~\ref{fig:spec}, in close on-sky separation to a nearby companion galaxy (the {\it `blueberry'}) offset by $\sim 0\farcs 5$ or 3 kpc at the same redshift, $z=5.124$.



\section{Analysis and results} \label{sec:res}

\subsection{The Banana and Blueberry}

The main galaxy \src\ has a puzzling banana-like morphology, characterized by a bright, central `knot' and more diffuse, extended emission as evidenced in Fig.~\ref{fig:spec}. The origin of the extended emission is not obvious. For instance, there is no evidence of a strong foreground lensing galaxy or cluster in the images, which could otherwise cause the observed shear of the source. In the center of the knot we find a strongly dominant blue emission, here characterized by the \jwst/NIRCam F115W filter, whereas the outskirts and the extended emission appear redder (in F356W and F444W). These gradients are commonly observed in high-redshift galaxies with \jwst\ \citep[e.g.,][]{Miller22,GimenezArteaga23}, hinting at younger or less dusty stellar populations in the centers. In contrast, the companion `blueberry' galaxy offset by $\sim 0\farcs 5$ corresponding to 3 kpc at $z=5.124$ shows a simple, 2D Gaussian-like profile with no hints of lensing-shear and with overall stronger continuum emission in the bluer bands. We will discuss the origin of the `blueberry' companion in more detail in Sect.~\ref{sec:popiii}. 

\subsection{Redshift and emission-line modeling}

From the extracted \jwst/NIRSpec Prism 1D spectrum, we identify several prominent rest-frame UV to optical nebular emission lines as marked in Fig.~\ref{fig:spec}. The broad Prism wavelength coverage allows us to constrain all lines from Ly$\alpha$ at 121 nm out to 900 nm at the source redshift. We model the continuum with a simple polynomial and the detected emission lines with Gaussian line profiles, tying the redshift $z_{\rm spec}$ and the intrinsic full-width-half-maximum (FWHM) across the transitions. The model is convolved by the \jwst/NIRSpec wavelength-dependent spectral resolution \citep{Jakobsen22}, here multiplied by a factor $1.3\times$ from the nominal to match the observed line widths across the spectrum \citep{DeGraaff24}. Consequently, we do not find evidence for lines intrinsically broader than the instrumental resolution ($\gtrsim 1000$\,km\,s$^{-1}$). The identified lines are marked Fig.~\ref{fig:spec} and their flux densities and rest-frame equivalent widths (EWs) are reported in Table~\ref{tab:lines}.   


We derive a spectroscopic redshift for \src\ of $z_{\rm spec} = 5.1240\pm 0.0002$, consistent with that reported from the automatic fitting in DJA. The galaxy is characterized by a relatively high ionization parameter, with an [\oiii]/[\oii] line ratio of $8.8\pm 1.1$. The Balmer H$\alpha$/H$\beta$ line ratio is consistent with the intrinsic ratio expected for the Case B recombination scenario at $T=2\times 10^{4}\,$K \citep{Osterbrock06}, implying minimal nebular dust attenuation.
Using the relevant set of strong-line diagnostics from \citet{Sanders24} based on the lines detected in the spectrum implies a relatively low metallicity, quantified in terms of the oxygen abundance as $12+\log({\rm O/H}) = 7.88\pm 0.07$ corresponding to $0.15\,Z/Z_{\odot}$ solar metallicity, typical for galaxies at $z\approx 5$ with stellar masses, $M_\star \approx 10^{9}\,M_\odot$ \citep[e.g.,][]{Heintz23,Nakajima23}. 

Based on the substantial EWs of the Balmer lines, for instance H$\alpha$ is measured to have ${\rm EW_{H\alpha}} = 734 \pm 33\,\AA$, we infer a nebular-to-stellar light fraction of $X_{\rm neb} \approx 10\%$ using the empirical relation described in \citet{Miranda25}. This is seemingly at odds with the prominent Balmer `jump' observed in the spectrum, implying a substantially higher nebular-to-stellar continuum fraction. We will consider the scenario of strong nebular continuum further in the sections below. 

\begin{table}
     \renewcommand{\arraystretch}{1.0} 
      \setlength{\tabcolsep}{5pt} 
            \caption{Detected emission-line transitions, flux densities, and rest-frame equivalent widths.}
        \begin{tabular}{ l c c } 
         \hline \hline    
            Transition & Line flux & Rest-frame EW \\
            & ($\times 10^{18}$ erg\,s$^{-1}$\,cm$^{-2}$) & ($\AA$) \\
            \hline 
            [O\,\textsc{ii}] $\lambda\lambda 3726,3729$ & $4.2 \pm 0.5$ & $35.9 \pm 4$ \\
            $[\rm{Ne\,\textsc{iii}}] $$\lambda 3869$ & $2.5 \pm 0.4$ & $24 \pm 5$\\ 
            H$\gamma$ & $2.8 \pm 0.3$ & $35.14 \pm 4$ \\
            H$\beta$ & $4.3 \pm 0.3$ & $137.3 \pm 13$ \\ 
            $[\rm{O\,\textsc{iii}}]$\,$\lambda\lambda 4959,5007 $ & $29.6 \pm 0.4$ & $946 \pm 46$\\
            He\,{\sc i} $\lambda 5876$ & $ 0.6 \pm 0.1$ & $15 \pm 4$ \\ 
            H$\alpha$ & $13.8 \pm 0.2$ & $734 \pm 33$ \\
            He\,{\sc i}/S\,{\sc i} & $0.5 \pm 0.2$ & $24 \pm 11$ \\ 
            \hline \hline  
        \end{tabular}  \\\\
     \label{tab:lines}
\end{table}

\subsection{Rest-frame UV properties}

The integrated rest-frame UV ($\sim 1500\,\AA$) flux density of \src implies a relatively bright continuum, with absolute UV magnitude $M_{\rm UV} = -20.95\pm 0.05$\,mag. This is among the top few percent most luminous sources at $z>5$ when comparing to the archival \jwst/NIRSpec Prism `PRIMAL' sample compiled by \citet{Heintz25}. Modeling the rest-frame UV spectral shape from $1250-2600\,\AA$ with a standard power-law, $F_{\lambda} \propto \lambda^{\beta_{\rm UV}}$, yields a relatively `red' continuum with $\beta_{\rm UV} = -1.68\pm 0.05$. This implies either a substantial amount of dust, $A_V\sim 0.5$\,mag, which is inconsistent with the low Balmer decrement, so more likely due to the rest-frame UV being dominated by the nebular continuum.

\begin{figure}
    \centering
    \includegraphics[width=9.2cm]{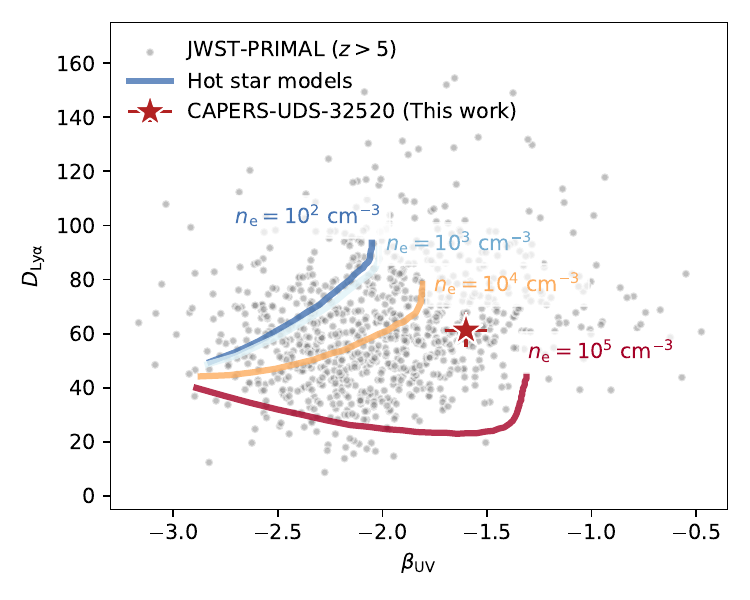}
    \caption{Ly$\alpha$ damping parameter, $D_{\rm Ly\alpha}$, as a function of the rest-frame UV spectral slope, $\beta_{\rm UV}$ for \src\ (red star symbol). For comparison, the grey dots show the full PRIMAL sample at $z>5$ \citep[from][]{Heintz25} and the colored tracks show predictions for hot metal-poor star models at various gas densities \citep[from][]{Katz25}.}
    \label{fig:betadlya}
\end{figure}

To quantify the physical conditions implied for the nebulae from the rest-frame UV continuum, we first consider strength of the UV rollover, quantified via the damping Ly$\alpha$ ($D_{\rm Ly\alpha}$) parameter as defined by \citet{Heintz25}. This basically measures the EW over the Ly$\alpha$ region from $1180$ to $1350\,\AA$ in the rest-frame. We measure $D_{\rm Ly\alpha} = 61.3\pm 6.2\,\AA$, implying either strong damped Ly$\alpha$ absorption (DLA) from \hi\ with column density $N_{\rm HI}\sim 10^{23}\,{\rm cm}^{-2}$ or a pronounced $2\gamma$ emission \citep{Katz25}. These scenarios will be discussed in more detail in Sect.~\ref{sec:uvroll} below. As a first test, we plot the measured UV characteristics, $\beta_{\rm UV}$ and $D_{\rm Ly\alpha}$, in Fig.~\ref{fig:betadlya}, compared to the full \jwst-PRIMAL sample \citep[][]{Heintz25} and to predictions for the parameter space covered by hot metal-poor star models for a range of \textbf{\textit{stellar}} temperatures $T\sim 10^{4}-10^{5}$\,K and densities $n_e\sim 10^2-10^{5}\,{\rm cm}^{-3}$ from \citet{Katz25}. Crucially, there is a unique combination between the $\beta_{\rm UV}$ and $D_{\rm Ly\alpha}$ for a given gas temperature and density predicted from the models. From the observed parameters for \src\ we infer substantial gas densities $n_{\rm e} \gtrsim 10^{4}\,{\rm cm}^{-3}$ and high temperatures, $T\gtrsim 8\times 10^{4}$\,K, based on this model. This also implies a substantial nebular-to-stellar continuum light-ratio of $\sim 50\%$ at 1500\,\AA\ and $\gtrsim 90\%$ at 3640\,\AA\ rest-frame \citep{Katz25}. 

\section{The origin of the strong UV-turnover} \label{sec:uvroll}

The most prominent characteristic of this galaxy is the strong UV rollover, in combination with the red UV spectral slope and the pronounced Balmer jump. To carefully examine the origin of the broad \lya\ damping wing and the overall spectral shape, we consider two distinct models: i) an underlying stellar continuum imprinted with a strong damped \lya\ absorber (DLA) due to dense \hi\ gas, see Sect.~\ref{ssec:DLA}, or ii) a rest-frame UV to optical continuum dominated by $2\gamma$ and free-bound nebular emission, see Sect.~\ref{ssec:neb}. For both scenarios, we mask out all the most prominent emission lines in the fitting, since both models are only sensitive to the continuum emission. 


\subsection{DLA model}\label{ssec:DLA}

To model the damped \lya\ wing, we adopt the method detailed in \citet[][Pollock et al., (subm.)]{Heintz24, Terp2024} to constrain $N_{\rm HI}$ in high-redshift galaxy spectra. 
We first assume an intrinsic rest-UV continuum described by a power law $F_\lambda = F_0\lambda^{\beta_{\rm UV}}$. The derived column density is typically insensitive to the exact underlying continuum, at least for strong \lya\ damping wings \citep[e.g.,][]{Heintz24}. The column density $N_{\rm HI}$ is derived by approximating the \lya\ absorption profile as a Voigt-Hjerting function, following the prescription from \citet{Tepper_Garc_a_2006}. 
We also include an additional component to represent the absorption from neutral hydrogen in the IGM due to the Gunn-Peterson effect, following the method described in \citet{MiraldaEscude98, Totani06}. We choose a fixed neutral hydrogen fraction of $x_{\rm HI}=10^{-3}$, a conservative choice given that the universe on average is expected to be completely ionized by $z=5.3$ \citep[e.g.][]{Bosman22}. We emphasize that any damping component will be dominated by DLA gas, especially in the low resolution Prism spectrum \citep[see e.g.][]{Huberty25, Heintz25, Mason26}.

\begin{figure}
    \centering
    \includegraphics[width=9.2cm]{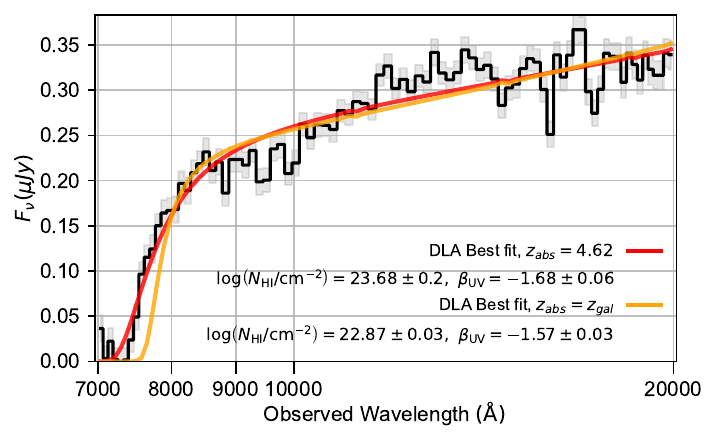}
    \caption{\jwst/NIRSpec Prism 1D spectrum, zoomed in on the rest-frame UV region. The best-fit DLA models are shown, assuming either a fixed DLA redshift to $z_{\rm gal}$ (yellow) or leaving the absorption redshift as a free parameter (red). The median and 16th to 84th percentiles on the output parameters from both models are summarized in the bottom-right legend. }
    \label{fig:dla_fit}
\end{figure}

\begin{figure*}[!t]
    \centering
    \includegraphics[width=18cm]{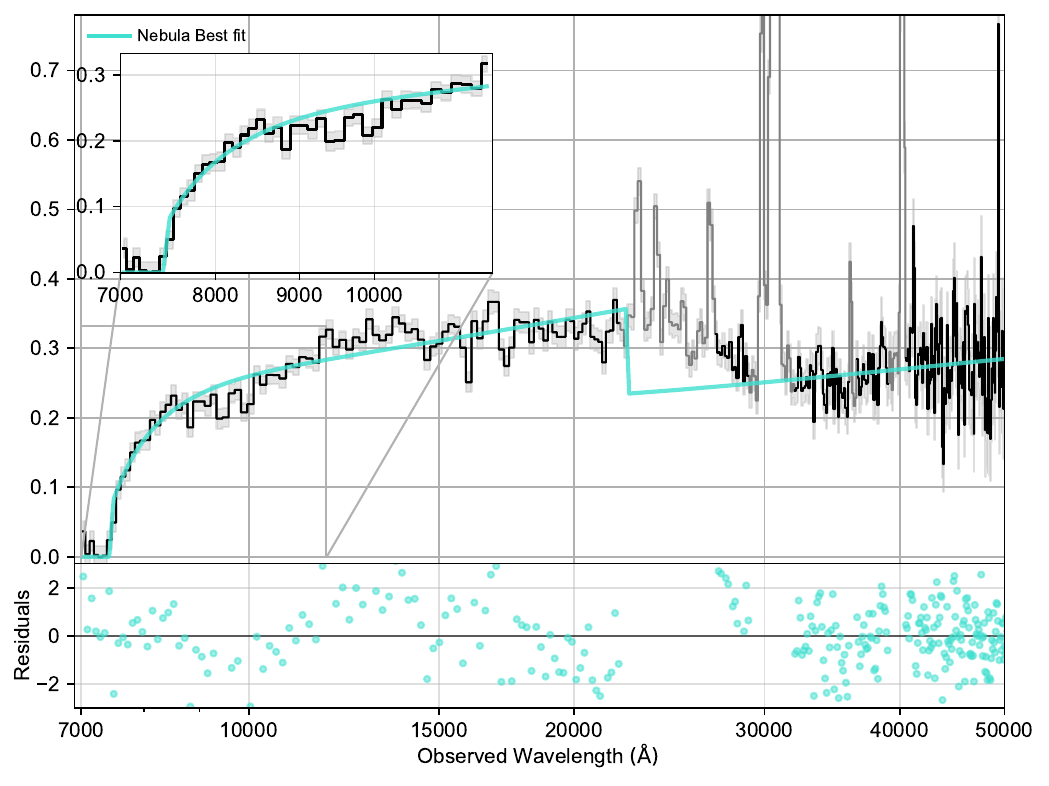}
    \caption{\jwst/NIRSpec Prism 1D spectrum of \src\ (black), overlaid the best-fit nebular continuum emission model from \texttt{PyNeb} (cyan). The model is only constrained by the continuum, with the excluded regions shown in grey. This nebular-dominated model yields physical conditions of $T = (5.3 \pm 0.2) \times 10^4$\,K and $n_e$ = $(5.4 \pm 0.8) \times 10^3\,{\rm cm}^{-3}$.}
    \label{fig:neb_fit}
\end{figure*}

When first considering a model with fixed absorption redshift to the systemic redshift of the galaxy, $z_{\rm abs} = 5.124$, we obtain a best-fit of $N_{\rm HI} = 10^{22.87\pm 0.03} \rm cm^{-2}$ as shown by the yellow model in Figure~\ref{fig:dla_fit}. This model visually does not provide a good match to the data.
Instead, leaving the absorption redshift as a free parameter, inspired by the galaxy and gas overdensity identified toward several background galaxy recovered by \citet{Terp2024,Heintz26}, we derive $z_{abs}=4.62$, and $N_{\rm HI} = 10^{23.68\pm 0.02} \rm cm^{-2}$. Although this model is an improved visual and statistically preferred fit (red. $\chi^2 =2.9$ compared to red. 
$\chi^2 =5.0$), we caution that there is no additional evidence of a neutral gas reservoir of this magnitude or galaxy overdensities located in front of the galaxy. The extreme $z_{\rm abs}$ and $N_{\rm HI}$ derived are likely artifacts of the modeling, necessary to shift the wing and reproduce the trough seen at $\sim7000\rm \AA$. The highly unusual shape of the turnover is thus unlikely to be fully explained by a \lya\ damping profile from local, dense \hi\ gas. Finally, we note that the flux deficit seen in the spectrum from the model around $0.9-1.0\mu$m likely arise from blends of several strong Si\,{\sc iv}\,$\lambda 1393,1402$ and C\,{\sc iv}\,$\lambda 1548, 1550$ absorption lines in the spectrum. Alternatively, this absorption `bump' would match the $2175\,\AA$ dust bump at the suggested foreground $z_{\rm abs} = 4.62$. We find this unlikely, however, given the rareness of this feature in typical high-redshift galaxies \citep[but see e.g.,][]{Witstok23,Ormerod25}.

\subsection{Nebular-dominated model} \label{ssec:neb}

To test the most likely alternative scenario, we attempt to model the turnover expected from a dominant nebular continuum spectrum. To model the nebular continuum, we use the \texttt{\_get\_continuum1} function from \texttt{PyNeb} \citep{PyNeb}. The built-in $T_{max}$ was increased from the nominal $10^4$\,K to $10^5$\,K to be able to fully reproduce the observed Balmer jump. 

With this alternative model, we are able to model and constrain the entire continuum spectrum, from $\sim0.7 - 5.3 \mu \rm m$. 
Our best-fit model implies an extreme nebular-dominated galaxy, with a gas temperature of $T = (5.3 \pm 0.2) \times 10^4$\,K and density $n_e$ = $(5.4 \pm 0.8) \times 10^3\,{\rm cm}^{-3}$, see Fig.~\ref{fig:neb_fit}. We note that this density is lower by a factor of $\approx 10$ to the one predicted from just the $D_{\rm Ly\alpha}$ and $\beta_{\rm UV}$ alone, suggesting that these two parameters still provide a too simple overview of the actual physical properties. The derived temperatures are substantially higher than seen in typical \hii\ region at high redshifts though the densities match those derived from high-ion transitions \citep[e.g.,][]{Topping25}. The strength of the Balmer jump is highly sensitive to the nebular temperature \citep[e.g.,][]{Trussler2025}, yielding relatively tight constraints on this parameter. 
We include an underlying power-law in the modeling, representing the potential contribution from the stellar population in the galaxy, finding that it can only contribute up to a few percent of the observed spectrum. As a consequence, the stellar UV luminosity is of the order $M_{\rm UV} = -17.7$\,mag, significantly less than the total observed value of $M_{\rm UV} = -20.95\pm 0.05$\,mag. 

\begin{figure*}
    \centering
    \includegraphics[width=0.8\linewidth]{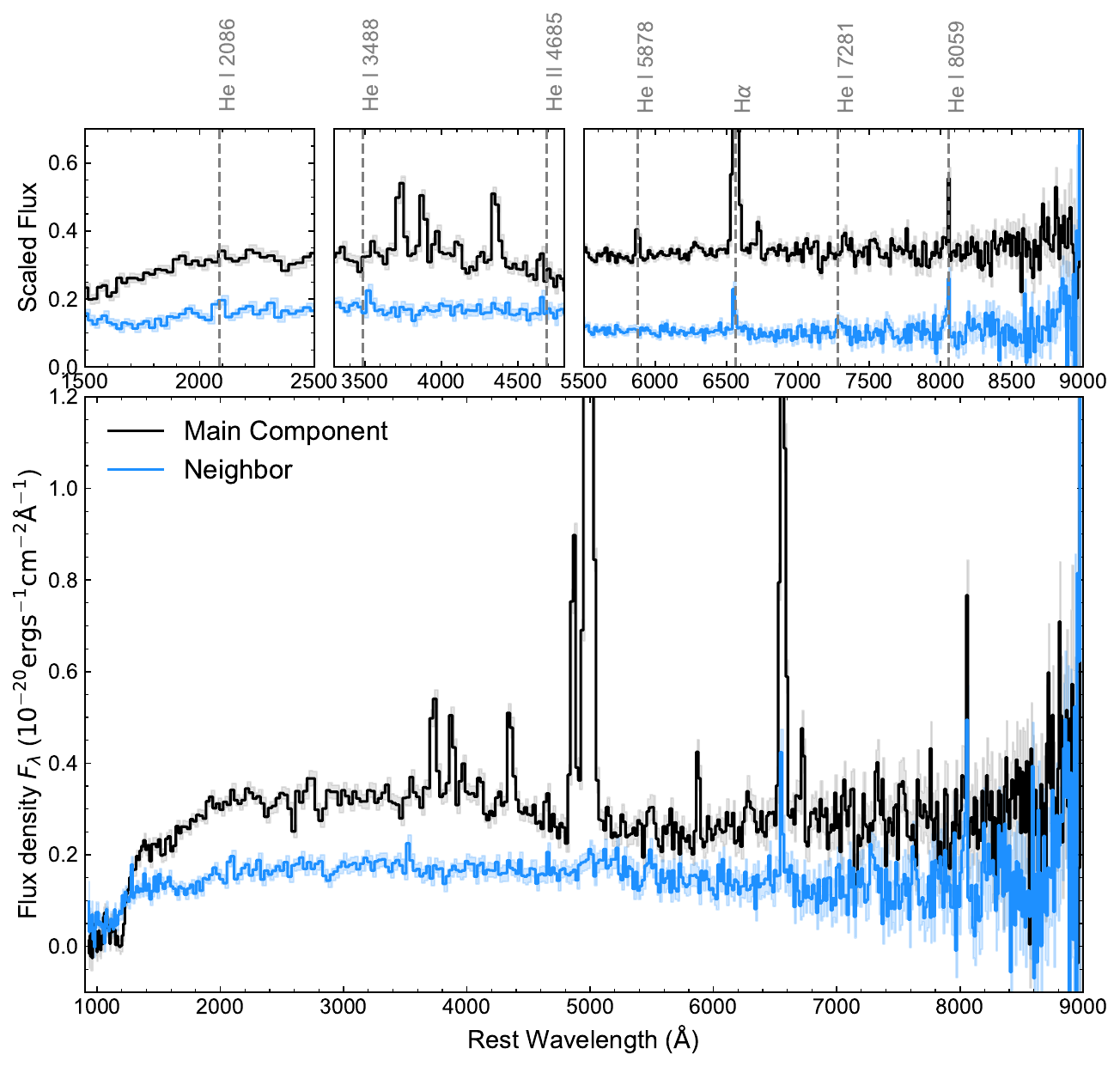}
    \caption{JWST/Prism spectra of the main galaxy (black) and the neighbor/``blueberry" (blue). Insets show zoom-ins on potentially weak Hydrogen and Helium emission lines in both components, with a lack of any metal lines in the companion.}
    \label{fig:popiii}
\end{figure*}

While this model provides a remarkable match to the data (red. $\chi^2 =2.3$) across the entire wavelength range from the rest-frame UV to optical, there is still an apparent flux deficit at $\sim 0.9-1.0\mu$m. This is expected if originating from a pseudo absorption-continuum from a blend of several high-ion absorption lines as described above. Further, the underlying continuum to the emission lines observed at $~2.3-2.8\mu$m appear higher than predicted from the otherwise sharp Balmer jump transition. It is not evident what the origin of this discrepancy is, but we hypothesize a potential flux excess due to multiple blends of emission features at those wavelengths. Finally, we highlight the  the lack of a strong \lya\ emission line in the spectrum, which are otherwise often observed in similar nebular-dominated galaxies \citep[e.g.,][]{Cameron2024_NDG,Katz25}. This could potentially be explained by an extra DLA component in the spectra, absorbing the \lya\ photons. At the derived spectral resolution we are, however, not able to accurately constrain this, finding an approximate upper bound of $N_{\rm HI} \lesssim 10^{21}\,{\rm cm^{-2}}$ when modeled together with the nebular-continuum model.





\section{Are PopIII stars driving the hot nebular continuum?} \label{sec:popiii}

Under a given set of ISM conditions, the strength of the nebular continuum scales with the strength of the ionizing field powering the emission.
Adopting standard stellar population models (including a Milky Way IMF), stellar populations with sufficiently young ages can power strong nebular continuum that becomes significant at rest-frame optical wavelengths, observable as a so-called Balmer jump \citep[e.g.,][]{Guseva2007}. 
However, these stellar populations will produce such a strong UV continuum that at $\lambda_{\rm rest}\approx1500$~\AA{} the nebular continuum should be completely sub-dominant.
For the nebular continuum to make a significant contribution to the emergent flux at UV wavelengths, a stellar population with ionizing photon production efficiency $\log (\xi_{\rm ion})\gtrsim 25.8$ is required. Such a high value of $\xi_{\rm ion}$ is only physically possible with a significant presence of very hot stars with stellar temperatures $T_{\rm eff}\gtrsim80,000$~K \citep{Katz25}.

This particular presence of very hot stars was invoked by \citet{Cameron2024_NDG} to explain the high inferred UV nebular fraction, with those authors highlighting $Z=0.07~Z_\odot$ Wolf-Rayet stars \citep{Todt2015}, binary stripped stars \citep{Gotberg2018}, or metal-free (Pop. III) stars \citep{Zackrisson2011, Larkin2023} as possible sources of the derived hot ionizing continuum.
Indeed, a nebular-dominated spectrum is a generic prediction for the emergent spectra Pop. III stellar populations \citep{Raiter2010, Trussler2023}.

The detection of oxygen and neon metal lines clearly demonstrates that the ISM of \src\ has been enriched with metals, with a metallicity of $12+\log({\rm O}/{\rm H})\approx 7.8-8.2$ derived from the measured line ratios.
For the observed emission to be powered by similarly metal-enriched stars, these would be required to be in a hot, evolved phase (e.g. Wolf-Rayet stars), perhaps also with a stellar IMF skewed towards higher mass (i.e. `top-heavy') to further enhance the contribution of the hotter stars \citep[e.g.,][]{Cameron2024_NDG, Cullen2025}.
Alternatively, we may be witnessing metal-enriched gas being photoionized by a cluster of Pop. III stars.

With this in mind, we revisit the `blueberry' companion, which serendipitously lies across the slit mask, allowing us to extract the spectrum separately to the main component, which can be compared in Figure~\ref{fig:popiii}. 
This spectrum shows clear H$\alpha$ emission at the same redshift as \src. We additionally identify potential He\,{\sc i} and He\,{\sc ii} emission lines, particularly He\,{\sc i}\,$\lambda8059$, He\,{\sc i}\,$\lambda7281$, He\,{\sc i}\,$\lambda5878$ (only in main component), He\,{\sc ii}\,$\lambda4685$, He\,{\sc i}\,$\lambda3488$, and He\,{\sc i}\,$\lambda2086$. 
We caution, however, that the photometric data for the blueberry from the {\tt Minerva} survey \citep[][]{Muzzin25} strongly favors a lower-redshift solution at $z\sim 0.8$ (C. Willott, priv. comm.). Consequently, this would explain the observed break at the same wavelength as \lya\ at $z=5.1$ to instead be from the 4000\,\AA\ Balmer break. The imposed H and He emission lines would then potentially extend from the main galaxy, whereas the nebular metal lines would be more centrally located in the \hii\ regions of \src. We argue that the presence of such an extended pristine gaseous region is also unlikely, and therefore keep both scenarios as likely interpretations. 

A key prediction for the survival of observable Pop. III galaxies to redshifts $z\sim5$ is exactly the presence of pristine low-mass halos, or smaller satellite galaxies \citep[e.g.][]{Zier25, Yajima23}. These systems can evolve slower, and maintain pristine gas for longer; avoiding chemical enrichment and metal pollution from supernovae in the larger companion galaxy. They may also continue to accrete gas from the surrounding circumgalactic medium (CGM). Pop. III remnants may therefore survive in satellite galaxies at fairly low redshifts, providing an opportunity to directly observe them with \jwst. 

Several observations of Pop III-like candidate systems have also previously been proposed in the literature, including the He {\sc ii}$\lambda1640$ clump detected 20~kpc offset from GN-z11 \citep{Maiolino24}, the magnified systems at $z=6.6$ \citep{Vanzella23, Nakajima25} and $z=6.5$ \citep{Fujimoto25} with bright Balmer lines but weak [\oiii] metal lines, an extremely metal-poor galaxy ($\sim1.6\%$ solar) at $z=8.27$ \citep{Cullen25}, and a potential hybrid Pop. II and Pop. III system with strong He {\sc ii} emission but no UV metal lines \citep{Wang24}. 
There is thus mounting evidence for the presence of these pristine, PopIII-like galaxies or halo systems around high-redshift galaxies uncovered by \jwst.

\section{Conclusions} \label{sec:conc}

In this work, we presented the discovery and characterization of a particularly strong nebular emission dominated galaxy at $z=5.124$ observed as part of the CAPERS survey, called \src. The galaxy was identified to be likely nebular-dominated in the rest-frame UV and optical based on the prominent Balmer `jump' and strong UV rollover near \lya\ observed in the \jwst/NIRSpec spectrum, in addition to several strong nebular emission lines. This galaxy was particularly interesting in connection with the recent literature on similar such galaxies \citep[e.g.,][]{Cameron2024_NDG,Katz25}, since its rest-frame UV properties indicated a particularly hot ($T\sim 5\times 10^4$\,K) and dense ($n_e > 10^{3}$\,cm$^{-3}$) gaseous region. These nebular-dominated galaxies are particularly interesting because their high gas temperatures imply exceptionally hot and potentially Pop. III stars powering their emission.

We found the galaxy to show a peculiar, banana-like morphology, with a bright central `knot' where the NIRSpec shutter was centered and an extended, curved emission in the north and south directions. This could hint at a potential disturbed or merger-like origin. Intriguingly, the \jwst/NIRCam imaging from the PRIMER survey of the field revealed a `blue' companion galaxy at potentially the same redshift as the main source, offset by 3\,kpc. We discussed the origin of this blue companion, the spectrum of which reveal no metal lines but several tentative hydrogen and helium lines. This indicated that the companion galaxy, if confirmed, is likely pristine. Such pristine low-mass halos are a key prediction for the survival of observable PopIII galaxies at $z\sim 5$. These results thus presented an independent validation of this physical connection.  

We analyzed the origin of the observed continuum and emission-line spectrum in detail, and found that it had a high ionization parameter with a [\oiii]/[\oii] line ratio of $8.8\pm 1.1$. The Balmer decrement indicated negligible amounts of dust and based on the [\oiii]/H$\beta$ ratio the galaxy is likely relatively metal-poor with 12+log(O/H) = $8.0\pm 0.2$, typical for galaxies with stellar masses of $\sim 10^9\,M_\odot$ at $z\approx 5$. We found that the model with a hot and dense nebula, with gas temperature $T = (5.3\pm 0.2)\times 10^{4}$\,K and $n_e = (5.4\pm 0.8)\times 10^{3}$\,cm$^{-3}$ was able to reproduce the entire rest-frame UV to optical continuum emission. 

This model further implied that the dominant fraction ($>95\%$) of the emitted light is originating from the nebular continuum. Consequently, the stellar light of this galaxy would only have an observed UV magnitude of $M_{\rm UV} = -17.7$\,mag, significantly less than the total observed value of $M_{\rm UV} = -20.95\pm 0.05$\,mag. This has strong implications for the presence of the apparent overabundant population of UV-bright galaxies at high redshifts, which might be artificially boosted if a substantial population of galaxies with strong nebular continuum exists. Gauging the impact of this population to the full underlying population of galaxies at cosmic dawn is thus essential to potentially alleviate some of the current tension in the field.

\begin{acknowledgements}
We would like to thank Harley Katz for valuable and insightful discussions during the initial stages of this work, and for sharing his code to model the nebular continuum. We also thank Chris Willott for kindly sharing the photometric data and pointing out the lower-redshift solution for the ``blueberry''. We finally express our strongest gratitude to the investigators on the CAPERS observing programs. The work presented here would not have been possible without their major efforts in designing and obtaining the observational data included in our work here. \\\\

The data products presented herein were retrieved from the Dawn JWST Archive (DJA). DJA is an initiative of the Cosmic Dawn Center, which is funded by the Danish National Research Foundation under grant DNRF140.
KEH acknowledges support from the Independent Research Fund Denmark (DFF) under grant 5251-00009B and co-funding by the European Union (ERC, HEAVYMETAL, 101071865). Views and opinions expressed are, however, those of the authors only and do not necessarily reflect those of the European Union or the European Research Council. Neither the European Union nor the granting authority can be held responsible for them. 
This work is based in part on observations made with the NASA/ESA/CSA James Webb Space Telescope. The data were obtained from the Mikulski Archive for Space Telescopes (MAST) at the Space Telescope Science Institute, which is operated by the Association of Universities for Research in Astronomy, Inc., under NASA contract NAS 5-03127 for JWST. \\\\

Software used in this work includes \textsc{matplotlib} \citep{Matplotlib}, \textsc{numpy} \citep{Numpy}, \textsc{astropy} \citep{Astropy}, and \textsc{scipy} \citep{scipy}. 
\end{acknowledgements}

%

\bibliographystyle{aa}
\bibliography{ref.bib}

@ARTICLE{Heintz25,
       author = {{Heintz}, K.~E. and {Brammer}, G.~B. and {Watson}, D. and {Oesch}, P.~A. and {Keating}, L.~C. and {Hayes}, M.~J. and {Abdurro'uf} and {Arellano-C{\'o}rdova}, K.~Z. and {Carnall}, A.~C. and {Christiansen}, C.~R. and {Cullen}, F. and {Dav{\'e}}, R. and {Dayal}, P. and {Ferrara}, A. and {Finlator}, K. and {Fynbo}, J.~P.~U. and {Flury}, S.~R. and {Gelli}, V. and {Gillman}, S. and {Gottumukkala}, R. and {Gould}, K. and {Greve}, T.~R. and {Hardin}, S.~E. and {Hsiao}, T.~Y. -Y. and {Hutter}, A. and {Jakobsson}, P. and {Killi}, M. and {Khosravaninezhad}, N. and {Laursen}, P. and {Lee}, M.~M. and {Magdis}, G.~E. and {Matthee}, J. and {Naidu}, R.~P. and {Narayanan}, D. and {Pollock}, C. and {Prescott}, M.~K.~M. and {Rusakov}, V. and {Shuntov}, M. and {Sneppen}, A. and {Smit}, R. and {Tanvir}, N.~R. and {Terp}, C. and {Toft}, S. and {Valentino}, F. and {Vijayan}, A.~P. and {Weaver}, J.~R. and {Wise}, J.~H. and {Witstok}, J.},
        title = "{The JWST-PRIMAL archival survey: A JWST/NIRSpec reference sample for the physical properties and Lyman-{\ensuremath{\alpha}} absorption and emission of {\ensuremath{\sim}}600 galaxies at z = 5.0 ‑ 13.4}",
      journal = {\aap},
         year = 2025,
        month = jan,
       volume = {693},
          eid = {A60},
        pages = {A60},
          doi = {10.1051/0004-6361/202450243},
archivePrefix = {arXiv},
       eprint = {2404.02211},
 primaryClass = {astro-ph.GA},
       adsurl = {https://ui.adsabs.harvard.edu/abs/2025A&A...693A..60H}
}

@article{Tepper_Garc_a_2006,
   title={Voigt profile fitting to quasar absorption lines: an analytic approximation to the Voigt-Hjerting function: A new method to compute Voigt profiles},
   volume={369},
   ISSN={0035-8711},
   url={http://dx.doi.org/10.1111/j.1365-2966.2006.10450.x},
   DOI={10.1111/j.1365-2966.2006.10450.x},
   number={4},
   journal={Monthly Notices of the Royal Astronomical Society},
   publisher={Oxford University Press (OUP)},
   author={Tepper García, Thorsten},
   year={2006},
   month=jun, pages={2025–2035} }

@MISC{Dickinson24,
       author = {{Dickinson}, Mark and {Amorin}, Ricardo and {Arrabal Haro}, Pablo and {Bagley}, Micaela and {Barro}, Guillermo and {Buat}, Veronique and {Burgarella}, Denis and {Calabro'}, Antonello and {Carnall}, Adam and {Casey}, Caitlin M. and {Chworowsky}, Katherine and {Cleri}, Nikko J. and {Cole}, Justin and {Cooper}, Michael and {Cullen}, Fergus and {Daddi}, Emanuele and {Donnan}, Callum and {Dunlop}, James S. and {Elbaz}, David and {Ferguson}, Henry C. and {Fernandez}, Vital and {Finkelstein}, Steven L. and {Fontana}, Adriano and {Fujimoto}, Seiji and {Giavalisco}, Mauro and {Hamilton}, Timothy S. and {Hathi}, Nimish P. and {Hirschmann}, Michaela and {Hutchison}, Taylor Alexandra and {Juneau}, Stephanie and {Jung}, Intae and {Kartaltepe}, Jeyhan and {Kocevski}, Dale D. and {Koekemoer}, Anton M. and {Larson}, Rebecca L. and {Long}, Arianna and {Lucas}, Ray A. and {Mascia}, Sara and {McGrath}, Elizabeth and {McLeod}, Derek and {McLure}, Ross and {Napolitano}, Lorenzo and {Papovich}, Casey and {Pentericci}, Laura and {Perez Gonzalez}, Pablo G. and {Simons}, Raymond and {Somerville}, Rachel S. and {Trump}, Jonathan R. and {Wang}, Xin and {Weiner}, Benjamin and {Wilkins}, Stephen Matthew and {Yung}, L.~Y. Aaron and {Zavala}, Jorge},
        title = "{The CANDELS-Area Prism Epoch of Reionization Survey (CAPERS)}",
 howpublished = {JWST Proposal. Cycle 3, ID. \#6368},
         year = 2024,
        month = mar,
        pages = {6368},
       adsurl = {https://ui.adsabs.harvard.edu/abs/2024jwst.prop.6368D}
}

@ARTICLE{DeGraaff24,
       author = {{de Graaff}, Anna and {Brammer}, Gabriel and {Weibel}, Andrea and {Lewis}, Zach and {Maseda}, Michael V. and {Oesch}, Pascal A. and {Bezanson}, Rachel and {Boogaard}, Leindert A. and {Cleri}, Nikko J. and {Cooper}, Olivia R. and {Gottumukkala}, Rashmi and {Greene}, Jenny E. and {Hirschmann}, Michaela and {Hviding}, Raphael E. and {Katz}, Harley and {Labb{\'e}}, Ivo and {Leja}, Joel and {Matthee}, Jorryt and {McConachie}, Ian and {Miller}, Tim B. and {Naidu}, Rohan P. and {Price}, Sedona H. and {Rix}, Hans-Walter and {Setton}, David J. and {Suess}, Katherine A. and {Wang}, Bingjie and {Whitaker}, Katherine E. and {Williams}, Christina C.},
        title = "{RUBIES: A complete census of the bright and red distant Universe with JWST/NIRSpec}",
      journal = {\aap},
         year = 2025,
        month = may,
       volume = {697},
          eid = {A189},
        pages = {A189},
          doi = {10.1051/0004-6361/202452186},
archivePrefix = {arXiv},
       eprint = {2409.05948},
 primaryClass = {astro-ph.GA},
       adsurl = {https://ui.adsabs.harvard.edu/abs/2025A&A...697A.189D}
}

@ARTICLE{Katz25,
       author = {{Katz}, Harley and {Cameron}, Alex J. and {Saxena}, Aayush and {Barrufet}, Laia and {Choustikov}, Nichloas and {Cleri}, Nikko J. and {de Graff}, Anna and {Ellis}, Richard S. and {Fosbury}, Robert A.~E. and {Heintz}, Kasper E. and {Maseda}, Michael and {Matthee}, Jorryt and {McConachie}, Ian and {Oesch}, Pascal A.},
        title = "{21 Balmer Jump Street: The Nebular Continuum at High Redshift and Implications for the Bright Galaxy Problem, UV Continuum Slopes, and Early Stellar Populations}",
      journal = {The Open Journal of Astrophysics},
         year = 2025,
        month = jul,
       volume = {8},
          eid = {104},
        pages = {104},
          doi = {10.33232/001c.142570},
archivePrefix = {arXiv},
       eprint = {2408.03189},
 primaryClass = {astro-ph.GA},
       adsurl = {https://ui.adsabs.harvard.edu/abs/2025OJAp....8E.104K}
}

@MISC{Dunlop21,
       author = {{Dunlop}, James S. and {Abraham}, Roberto G. and {Ashby}, Matthew L.~N. and {Bagley}, Micaela and {Best}, Philip N. and {Bongiorno}, Angela and {Bouwens}, Rychard and {Bowler}, Rebecca A.~A. and {Brammer}, Gabriel and {Bremer}, Malcolm and {Calabro'}, Antonello and {Carnall}, Adam and {Castellano}, Marco and {Cirasuolo}, Michele and {Conselice}, Christopher and {Cullen}, Fergus and {Dave}, Romeel and {Dayal}, Pratika and {Dekel}, Avishai and {Dickinson}, Mark and {Duncan}, Kenneth James and {Elbaz}, David and {Ellis}, Richard S. and {Ferguson}, Harry C. and {Ferrara}, Andrea and {Finkelstein}, Steven L. and {Fontana}, Adriano and {Furlanetto}, Steven and {Fynbo}, Johan P.~U. and {Gallerani}, Simona and {Gardner}, Jonathan P. and {Giavalisco}, Mauro and {Grazian}, Andrea and {Grogin}, Norman and {Harikane}, Yuichi and {Hopkins}, Philip F. and {Ilbert}, Olivier and {Illingworth}, Garth D. and {Juneau}, Stephanie and {Jung}, Intae and {Kartaltepe}, Jeyhan and {Kassin}, Susan and {Kauffmann}, Olivier Benjamin and {Khochfar}, Sadegh and {Kirkpatrick}, Allison and {Kocevski}, Dale D. and {Koekemoer}, Anton M. and {Labbe}, Ivo and {Laporte}, Nicolas and {Larson}, Rebecca L. and {Lucas}, Ray A. and {Magee}, Daniel K. and {Mason}, Charlotte and {McCracken}, Henry Joy and {McLeod}, Derek and {McLure}, Ross and {Merlin}, Emiliano and {Mesinger}, Andrei and {Milvang-Jensen}, Bo and {Newman}, Jeffrey Allen and {Oesch}, Pascal and {Ouchi}, Masami and {Pacifici}, Camilla and {Papovich}, Casey and {Peacock}, John and {Peeples}, Molly and {Pentericci}, Laura and {Perez-Gonzalez}, Pablo G. and {Pirzkal}, Norbert and {Pope}, Alexandra and {Pye}, John P. and {Reddy}, Naveen A. and {Robertson}, Brant and {Salvato}, Mara and {Santini}, Paola and {Schaerer}, Daniel and {Shapley}, Alice E. and {Simons}, Raymond and {Smit}, Renske and {Smith}, Britton D. and {Snyder}, Greg and {Somerville}, Rachel S. and {Stanway}, Elizabeth R. and {Stefanon}, Mauro and {Tasca}, Lidia and {Tikkanen}, Tuomo and {Tresse}, Laurence and {Trump}, Jonathan R. and {Whitaker}, Katherine E. and {Wilkins}, Stephen Matthew and {Wright}, Gillian and {Wyithe}, J. Stuart B. and {van Dokkum}, Pieter and {van der Werf}, Paul},
        title = "{PRIMER: Public Release IMaging for Extragalactic Research}",
 howpublished = {JWST Proposal. Cycle 1, ID. \#1837},
         year = 2021,
        month = mar,
        pages = {1837},
       adsurl = {https://ui.adsabs.harvard.edu/abs/2021jwst.prop.1837D}
}

@ARTICLE{Miller22,
       author = {{Miller}, Tim B. and {Whitaker}, Katherine E. and {Nelson}, Erica J. and {van Dokkum}, Pieter and {Bezanson}, Rachel and {Brammer}, Gabriel and {Heintz}, Kasper E. and {Leja}, Joel and {Suess}, Katherine A. and {Weaver}, John R.},
        title = "{Early JWST Imaging Reveals Strong Optical and NIR Color Gradients in Galaxies at z   2 Driven Mostly by Dust}",
      journal = {\apjl},
         year = 2022,
        month = dec,
       volume = {941},
       number = {2},
          eid = {L37},
        pages = {L37},
          doi = {10.3847/2041-8213/aca675},
archivePrefix = {arXiv},
       eprint = {2209.12954},
 primaryClass = {astro-ph.GA},
       adsurl = {https://ui.adsabs.harvard.edu/abs/2022ApJ...941L..37M}
}

@ARTICLE{GimenezArteaga23,
       author = {{Gim{\'e}nez-Arteaga}, Clara and {Oesch}, Pascal A. and {Brammer}, Gabriel B. and {Valentino}, Francesco and {Mason}, Charlotte A. and {Weibel}, Andrea and {Barrufet}, Laia and {Fujimoto}, Seiji and {Heintz}, Kasper E. and {Nelson}, Erica J. and {Strait}, Victoria B. and {Suess}, Katherine A. and {Gibson}, Justus},
        title = "{Spatially Resolved Properties of Galaxies at 5 < z < 9 in the SMACS 0723 JWST ERO Field}",
      journal = {\apj},
         year = 2023,
        month = may,
       volume = {948},
       number = {2},
          eid = {126},
        pages = {126},
          doi = {10.3847/1538-4357/acc5ea},
archivePrefix = {arXiv},
       eprint = {2212.08670},
 primaryClass = {astro-ph.GA},
       adsurl = {https://ui.adsabs.harvard.edu/abs/2023ApJ...948..126G}
}

@BOOK{Osterbrock06,
       author = {{Osterbrock}, Donald E. and {Ferland}, Gary J.},
        title = "{Astrophysics of gaseous nebulae and active galactic nuclei}",
         year = 2006,
       adsurl = {https://ui.adsabs.harvard.edu/abs/2006agna.book.....O}
}

@ARTICLE{Sanders24,
       author = {{Sanders}, Ryan L. and {Shapley}, Alice E. and {Topping}, Michael W. and {Reddy}, Naveen A. and {Brammer}, Gabriel B.},
        title = "{Direct T $_{e}$-based Metallicities of z = 2{\textendash}9 Galaxies with JWST/NIRSpec: Empirical Metallicity Calibrations Applicable from Reionization to Cosmic Noon}",
      journal = {\apj},
         year = 2024,
        month = feb,
       volume = {962},
       number = {1},
          eid = {24},
        pages = {24},
          doi = {10.3847/1538-4357/ad15fc},
archivePrefix = {arXiv},
       eprint = {2303.08149},
 primaryClass = {astro-ph.GA},
       adsurl = {https://ui.adsabs.harvard.edu/abs/2024ApJ...962...24S}
}

@ARTICLE{Nakajima23,
       author = {{Nakajima}, Kimihiko and {Ouchi}, Masami and {Isobe}, Yuki and {Harikane}, Yuichi and {Zhang}, Yechi and {Ono}, Yoshiaki and {Umeda}, Hiroya and {Oguri}, Masamune},
        title = "{JWST Census for the Mass-Metallicity Star Formation Relations at z = 4-10 with Self-consistent Flux Calibration and Proper Metallicity Calibrators}",
      journal = {\apjs},
         year = 2023,
        month = dec,
       volume = {269},
       number = {2},
          eid = {33},
        pages = {33},
          doi = {10.3847/1538-4365/acd556},
archivePrefix = {arXiv},
       eprint = {2301.12825},
 primaryClass = {astro-ph.GA},
       adsurl = {https://ui.adsabs.harvard.edu/abs/2023ApJS..269...33N}
}

@ARTICLE{Heintz23,
       author = {{Heintz}, Kasper E. and {Brammer}, Gabriel B. and {Gim{\'e}nez-Arteaga}, Clara and {Strait}, Victoria B. and {Lagos}, Claudia del P. and {Vijayan}, Aswin P. and {Matthee}, Jorryt and {Watson}, Darach and {Mason}, Charlotte A. and {Hutter}, Anne and {Toft}, Sune and {Fynbo}, Johan P.~U. and {Oesch}, Pascal A.},
        title = "{Dilution of chemical enrichment in galaxies 600 Myr after the Big Bang}",
      journal = {Nature Astronomy},
         year = 2023,
        month = dec,
       volume = {7},
        pages = {1517-1524},
          doi = {10.1038/s41550-023-02078-7},
archivePrefix = {arXiv},
       eprint = {2212.02890},
 primaryClass = {astro-ph.GA},
       adsurl = {https://ui.adsabs.harvard.edu/abs/2023NatAs...7.1517H}
}

@ARTICLE{Gardner06,
       author = {{Gardner}, Jonathan P. and {Mather}, John C. and {Clampin}, Mark and {Doyon}, Rene and {Greenhouse}, Matthew A. and {Hammel}, Heidi B. and {Hutchings}, John B. and {Jakobsen}, Peter and {Lilly}, Simon J. and {Long}, Knox S. and {Lunine}, Jonathan I. and {McCaughrean}, Mark J. and {Mountain}, Matt and {Nella}, John and {Rieke}, George H. and {Rieke}, Marcia J. and {Rix}, Hans-Walter and {Smith}, Eric P. and {Sonneborn}, George and {Stiavelli}, Massimo and {Stockman}, H.~S. and {Windhorst}, Rogier A. and {Wright}, Gillian S.},
        title = "{The James Webb Space Telescope}",
      journal = {\ssr},
         year = 2006,
        month = apr,
       volume = {123},
       number = {4},
        pages = {485-606},
          doi = {10.1007/s11214-006-8315-7},
archivePrefix = {arXiv},
       eprint = {astro-ph/0606175},
 primaryClass = {astro-ph},
       adsurl = {https://ui.adsabs.harvard.edu/abs/2006SSRv..123..485G}
}

@ARTICLE{Heintz24,
       author = {{Heintz}, Kasper E. and {Watson}, Darach and {Brammer}, Gabriel and {Vejlgaard}, Simone and {Hutter}, Anne and {Strait}, Victoria B. and {Matthee}, Jorryt and {Oesch}, Pascal A. and {Jakobsson}, P{\'a}ll and {Tanvir}, Nial R. and {Laursen}, Peter and {Naidu}, Rohan P. and {Mason}, Charlotte A. and {Killi}, Meghana and {Jung}, Intae and {Hsiao}, Tiger Yu-Yang and {Abdurro'uf} and {Coe}, Dan and {Arrabal Haro}, Pablo and {Finkelstein}, Steven L. and {Toft}, Sune},
        title = "{Strong damped Lyman-{\ensuremath{\alpha}} absorption in young star-forming galaxies at redshifts 9 to 11}",
      journal = {Science},
         year = 2024,
        month = may,
       volume = {384},
       number = {6698},
        pages = {890-894},
          doi = {10.1126/science.adj0343},
archivePrefix = {arXiv},
       eprint = {2306.00647},
 primaryClass = {astro-ph.GA},
       adsurl = {https://ui.adsabs.harvard.edu/abs/2024Sci...384..890H}
}

@ARTICLE{Umeda24,
       author = {{Umeda}, Hiroya and {Ouchi}, Masami and {Nakajima}, Kimihiko and {Harikane}, Yuichi and {Ono}, Yoshiaki and {Xu}, Yi and {Isobe}, Yuki and {Zhang}, Yechi},
        title = "{JWST Measurements of Neutral Hydrogen Fractions and Ionized Bubble Sizes at z = 7{\textendash}12 Obtained with Ly{\ensuremath{\alpha}} Damping Wing Absorptions in 27 Bright Continuum Galaxies}",
      journal = {\apj},
         year = 2024,
        month = aug,
       volume = {971},
       number = {2},
          eid = {124},
        pages = {124},
          doi = {10.3847/1538-4357/ad554e},
archivePrefix = {arXiv},
       eprint = {2306.00487},
 primaryClass = {astro-ph.GA},
       adsurl = {https://ui.adsabs.harvard.edu/abs/2024ApJ...971..124U}
}

@ARTICLE{Hainline24,
       author = {{Hainline}, Kevin N. and {D'Eugenio}, Francesco and {Jakobsen}, Peter and {Chevallard}, Jacopo and {Carniani}, Stefano and {Witstok}, Joris and {Ji}, Zhiyuan and {Curtis-Lake}, Emma and {Johnson}, Benjamin D. and {Robertson}, Brant and {Tacchella}, Sandro and {Curti}, Mirko and {Charlot}, Stephane and {Helton}, Jakob M. and {Arribas}, Santiago and {Bhatawdekar}, Rachana and {Bunker}, Andrew J. and {Cameron}, Alex J. and {Egami}, Eiichi and {Eisenstein}, Daniel J. and {Hausen}, Ryan and {Kumari}, Nimisha and {Maiolino}, Roberto and {P{\'e}rez-Gonz{\'a}lez}, Pablo G. and {Rieke}, Marcia and {Saxena}, Aayush and {Scholtz}, Jan and {Smit}, Renske and {Sun}, Fengwu and {Williams}, Christina C. and {Willmer}, Christopher N.~A. and {Willott}, Chris},
        title = "{Searching for Emission Lines at z > 11: The Role of Damped Ly{\ensuremath{\alpha}} and Hints About the Escape of Ionizing Photons}",
      journal = {\apj},
         year = 2024,
        month = dec,
       volume = {976},
       number = {2},
          eid = {160},
        pages = {160},
          doi = {10.3847/1538-4357/ad8447},
archivePrefix = {arXiv},
       eprint = {2404.04325},
 primaryClass = {astro-ph.GA},
       adsurl = {https://ui.adsabs.harvard.edu/abs/2024ApJ...976..160H}
}

@ARTICLE{Witstok25,
       author = {{Witstok}, Joris and {Jakobsen}, Peter and {Maiolino}, Roberto and {Helton}, Jakob M. and {Johnson}, Benjamin D. and {Robertson}, Brant E. and {Tacchella}, Sandro and {Cameron}, Alex J. and {Smit}, Renske and {Bunker}, Andrew J. and {Saxena}, Aayush and {Sun}, Fengwu and {Alberts}, Stacey and {Arribas}, Santiago and {Baker}, William M. and {Bhatawdekar}, Rachana and {Boyett}, Kristan and {Cargile}, Phillip A. and {Carniani}, Stefano and {Charlot}, St{\'e}phane and {Chevallard}, Jacopo and {Curti}, Mirko and {Curtis-Lake}, Emma and {D'Eugenio}, Francesco and {Eisenstein}, Daniel J. and {Hainline}, Kevin N. and {Jones}, Gareth C. and {Kumari}, Nimisha and {Maseda}, Michael V. and {P{\'e}rez-Gonz{\'a}lez}, Pablo G. and {Rinaldi}, Pierluigi and {Scholtz}, Jan and {{\"U}bler}, Hannah and {Williams}, Christina C. and {Willmer}, Christopher N.~A. and {Willott}, Chris and {Zhu}, Yongda},
        title = "{Witnessing the onset of reionization through Lyman-{\ensuremath{\alpha}} emission at redshift 13}",
      journal = {\nat},
         year = 2025,
        month = mar,
       volume = {639},
       number = {8056},
        pages = {897-901},
          doi = {10.1038/s41586-025-08779-5},
archivePrefix = {arXiv},
       eprint = {2408.16608},
 primaryClass = {astro-ph.GA},
       adsurl = {https://ui.adsabs.harvard.edu/abs/2025Natur.639..897W}
}

@ARTICLE{Planck18,
       author = {{Planck Collaboration} and {Aghanim}, N. and {Akrami}, Y. and {Ashdown}, M. and {Aumont}, J. and {Baccigalupi}, C. and {Ballardini}, M. and {Banday}, A.~J. and {Barreiro}, R.~B. and {Bartolo}, N. and {Basak}, S. and {Battye}, R. and {Benabed}, K. and {Bernard}, J.-P. and {Bersanelli}, M. and {Bielewicz}, P. and {Bock}, J.~J. and {Bond}, J.~R. and {Borrill}, J. and {Bouchet}, F.~R. and {Boulanger}, F. and {Bucher}, M. and {Burigana}, C. and {Butler}, R.~C. and {Calabrese}, E. and {Cardoso}, J.-F. and {Carron}, J. and {Challinor}, A. and {Chiang}, H.~C. and {Chluba}, J. and {Colombo}, L.~P.~L. and {Combet}, C. and {Contreras}, D. and {Crill}, B.~P. and {Cuttaia}, F. and {de Bernardis}, P. and {de Zotti}, G. and {Delabrouille}, J. and {Delouis}, J.-M. and {Di Valentino}, E. and {Diego}, J.~M. and {Dor{\'e}}, O. and {Douspis}, M. and {Ducout}, A. and {Dupac}, X. and {Dusini}, S. and {Efstathiou}, G. and {Elsner}, F. and {En{\ss}lin}, T.~A. and {Eriksen}, H.~K. and {Fantaye}, Y. and {Farhang}, M. and {Fergusson}, J. and {Fernandez-Cobos}, R. and {Finelli}, F. and {Forastieri}, F. and {Frailis}, M. and {Fraisse}, A.~A. and {Franceschi}, E. and {Frolov}, A. and {Galeotta}, S. and {Galli}, S. and {Ganga}, K. and {G{\'e}nova-Santos}, R.~T. and {Gerbino}, M. and {Ghosh}, T. and {Gonz{\'a}lez-Nuevo}, J. and {G{\'o}rski}, K.~M. and {Gratton}, S. and {Gruppuso}, A. and {Gudmundsson}, J.~E. and {Hamann}, J. and {Handley}, W. and {Hansen}, F.~K. and {Herranz}, D. and {Hildebrandt}, S.~R. and {Hivon}, E. and {Huang}, Z. and {Jaffe}, A.~H. and {Jones}, W.~C. and {Karakci}, A. and {Keih{\"a}nen}, E. and {Keskitalo}, R. and {Kiiveri}, K. and {Kim}, J. and {Kisner}, T.~S. and {Knox}, L. and {Krachmalnicoff}, N. and {Kunz}, M. and {Kurki-Suonio}, H. and {Lagache}, G. and {Lamarre}, J.-M. and {Lasenby}, A. and {Lattanzi}, M. and {Lawrence}, C.~R. and {Le Jeune}, M. and {Lemos}, P. and {Lesgourgues}, J. and {Levrier}, F. and {Lewis}, A. and {Liguori}, M. and {Lilje}, P.~B. and {Lilley}, M. and {Lindholm}, V. and {L{\'o}pez-Caniego}, M. and {Lubin}, P.~M. and {Ma}, Y.-Z. and {Mac{\'\i}as-P{\'e}rez}, J.~F. and {Maggio}, G. and {Maino}, D. and {Mandolesi}, N. and {Mangilli}, A. and {Marcos-Caballero}, A. and {Maris}, M. and {Martin}, P.~G. and {Martinelli}, M. and {Mart{\'\i}nez-Gonz{\'a}lez}, E. and {Matarrese}, S. and {Mauri}, N. and {McEwen}, J.~D. and {Meinhold}, P.~R. and {Melchiorri}, A. and {Mennella}, A. and {Migliaccio}, M. and {Millea}, M. and {Mitra}, S. and {Miville-Desch{\^e}nes}, M.-A. and {Molinari}, D. and {Montier}, L. and {Morgante}, G. and {Moss}, A. and {Natoli}, P. and {N{\o}rgaard-Nielsen}, H.~U. and {Pagano}, L. and {Paoletti}, D. and {Partridge}, B. and {Patanchon}, G. and {Peiris}, H.~V. and {Perrotta}, F. and {Pettorino}, V. and {Piacentini}, F. and {Polastri}, L. and {Polenta}, G. and {Puget}, J.-L. and {Rachen}, J.~P. and {Reinecke}, M. and {Remazeilles}, M. and {Renzi}, A. and {Rocha}, G. and {Rosset}, C. and {Roudier}, G. and {Rubi{\~n}o-Mart{\'\i}n}, J.~A. and {Ruiz-Granados}, B. and {Salvati}, L. and {Sandri}, M. and {Savelainen}, M. and {Scott}, D. and {Shellard}, E.~P.~S. and {Sirignano}, C. and {Sirri}, G. and {Spencer}, L.~D. and {Sunyaev}, R. and {Suur-Uski}, A.-S. and {Tauber}, J.~A. and {Tavagnacco}, D. and {Tenti}, M. and {Toffolatti}, L. and {Tomasi}, M. and {Trombetti}, T. and {Valenziano}, L. and {Valiviita}, J. and {Van Tent}, B. and {Vibert}, L. and {Vielva}, P. and {Villa}, F. and {Vittorio}, N. and {Wandelt}, B.~D. and {Wehus}, I.~K. and {White}, M. and {White}, S.~D.~M. and {Zacchei}, A. and {Zonca}, A.},
        title = "{Planck 2018 results. VI. Cosmological parameters}",
      journal = {\aap},
         year = 2020,
        month = sep,
       volume = {641},
          eid = {A6},
        pages = {A6},
          doi = {10.1051/0004-6361/201833910},
archivePrefix = {arXiv},
       eprint = {1807.06209},
 primaryClass = {astro-ph.CO},
       adsurl = {https://ui.adsabs.harvard.edu/abs/2020A&A...641A...6P}
}

@ARTICLE{Astropy,
       author = {{Astropy Collaboration} and {Robitaille}, Thomas P. and {Tollerud}, Erik J. and {Greenfield}, Perry and {Droettboom}, Michael and {Bray}, Erik and {Aldcroft}, Tom and {Davis}, Matt and {Ginsburg}, Adam and {Price-Whelan}, Adrian M. and {Kerzendorf}, Wolfgang E. and {Conley}, Alexander and {Crighton}, Neil and {Barbary}, Kyle and {Muna}, Demitri and {Ferguson}, Henry and {Grollier}, Fr{\'e}d{\'e}ric and {Parikh}, Madhura M. and {Nair}, Prasanth H. and {Unther}, Hans M. and {Deil}, Christoph and {Woillez}, Julien and {Conseil}, Simon and {Kramer}, Roban and {Turner}, James E.~H. and {Singer}, Leo and {Fox}, Ryan and {Weaver}, Benjamin A. and {Zabalza}, Victor and {Edwards}, Zachary I. and {Azalee Bostroem}, K. and {Burke}, D.~J. and {Casey}, Andrew R. and {Crawford}, Steven M. and {Dencheva}, Nadia and {Ely}, Justin and {Jenness}, Tim and {Labrie}, Kathleen and {Lim}, Pey Lian and {Pierfederici}, Francesco and {Pontzen}, Andrew and {Ptak}, Andy and {Refsdal}, Brian and {Servillat}, Mathieu and {Streicher}, Ole},
        title = "{Astropy: A community Python package for astronomy}",
      journal = {\aap},
         year = 2013,
        month = oct,
       volume = {558},
          eid = {A33},
        pages = {A33},
          doi = {10.1051/0004-6361/201322068},
archivePrefix = {arXiv},
       eprint = {1307.6212},
 primaryClass = {astro-ph.IM},
       adsurl = {https://ui.adsabs.harvard.edu/abs/2013A&A...558A..33A}
}

@ARTICLE{Mason26,
       author = {{Mason}, Charlotte A. and {Chen}, Zuyi and {Stark}, Daniel P. and {Yi Lu}, Ting and {Topping}, Michael and {Tang}, Mengtao},
        title = "{Constraints on the z {\ensuremath{\sim}} 6{\ensuremath{-}}13 intergalactic medium from JWST spectroscopy of Lyman-alpha damping wings in galaxies}",
      journal = {\aap},
         year = 2026,
        month = jan,
       volume = {705},
          eid = {A114},
        pages = {A114},
          doi = {10.1051/0004-6361/202553820},
archivePrefix = {arXiv},
       eprint = {2501.11702},
 primaryClass = {astro-ph.GA},
       adsurl = {https://ui.adsabs.harvard.edu/abs/2026A&A...705A.114M}
}

@ARTICLE{Huberty25,
       author = {{Huberty}, Mason and {Scarlata}, Claudia and {Hayes}, Matthew J. and {Gazagnes}, Simon},
        title = "{The Pitfalls of Using Ly{\ensuremath{\alpha}} Damping Wings in High-z Galaxy Spectra to Measure the Intergalactic Neutral Hydrogen Fraction}",
      journal = {\apj},
         year = 2025,
        month = jul,
       volume = {987},
       number = {1},
          eid = {82},
        pages = {82},
          doi = {10.3847/1538-4357/add5e7},
archivePrefix = {arXiv},
       eprint = {2501.13899},
 primaryClass = {astro-ph.GA},
       adsurl = {https://ui.adsabs.harvard.edu/abs/2025ApJ...987...82H}
}

@ARTICLE{Heintz26,
       author = {{Heintz}, Kasper E. and {Bennett}, Jake S. and {Oesch}, Pascal A. and {Sneppen}, Albert and {Rennehan}, Douglas and {Pollock}, Clara L. and {Witstok}, Joris and {Smit}, Renske and {Vejlgaard}, Simone and {Terp}, Chamilla and {Koca}, Umran S. and {Brammer}, Gabriel B. and {Finlator}, Kristian and {Hayes}, Matthew J. and {Sijacki}, Debora and {Naidu}, Rohan P. and {Matthee}, Jorryt and {Valentino}, Francesco and {Tanvir}, Nial R. and {Jakobsson}, P{\'a}ll and {Laursen}, Peter and {Watson}, Darach J. and {Dav{\'e}}, Romeel and {Keating}, Laura C. and {Covelo-Paz}, Alba},
        title = "{A dense web of neutral gas in a galaxy proto-cluster post-reionization}",
      journal = {Nature Astronomy},
         year = 2026,
        month = jan,
          doi = {10.1038/s41550-025-02745-x},
       adsurl = {https://ui.adsabs.harvard.edu/abs/2026NatAs.tmp....4H}
}

@ARTICLE{Bosman22,
       author = {{Bosman}, Sarah E.~I. and {Davies}, Frederick B. and {Becker}, George D. and {Keating}, Laura C. and {Davies}, Rebecca L. and {Zhu}, Yongda and {Eilers}, Anna-Christina and {D'Odorico}, Valentina and {Bian}, Fuyan and {Bischetti}, Manuela and {Cristiani}, Stefano V. and {Fan}, Xiaohui and {Farina}, Emanuele P. and {Haehnelt}, Martin G. and {Hennawi}, Joseph F. and {Kulkarni}, Girish and {Mesinger}, Andrei and {Meyer}, Romain A. and {Onoue}, Masafusa and {Pallottini}, Andrea and {Qin}, Yuxiang and {Ryan-Weber}, Emma and {Schindler}, Jan-Torge and {Walter}, Fabian and {Wang}, Feige and {Yang}, Jinyi},
        title = "{Hydrogen reionization ends by z = 5.3: Lyman-{\ensuremath{\alpha}} optical depth measured by the XQR-30 sample}",
      journal = {\mnras},
         year = 2022,
        month = jul,
       volume = {514},
       number = {1},
        pages = {55-76},
          doi = {10.1093/mnras/stac1046},
archivePrefix = {arXiv},
       eprint = {2108.03699},
 primaryClass = {astro-ph.CO},
       adsurl = {https://ui.adsabs.harvard.edu/abs/2022MNRAS.514...55B}
}

@ARTICLE{MiraldaEscude98,
       author = {{Miralda-Escud{\'e}}, Jordi},
        title = "{Reionization of the Intergalactic Medium and the Damping Wing of the Gunn-Peterson Trough}",
      journal = {\apj},
         year = 1998,
        month = jul,
       volume = {501},
       number = {1},
        pages = {15-22},
          doi = {10.1086/305799},
archivePrefix = {arXiv},
       eprint = {astro-ph/9708253},
 primaryClass = {astro-ph},
       adsurl = {https://ui.adsabs.harvard.edu/abs/1998ApJ...501...15M}
}

@ARTICLE{Totani06,
       author = {{Totani}, Tomonori and {Kawai}, Nobuyuki and {Kosugi}, George and {Aoki}, Kentaro and {Yamada}, Toru and {Iye}, Masanori and {Ohta}, Kouji and {Hattori}, Takashi},
        title = "{Implications for Cosmic Reionization from the Optical Afterglow Spectrum of the Gamma-Ray Burst 050904 at z = 6.3$^{*}$}",
      journal = {\pasj},
         year = 2006,
        month = jun,
       volume = {58},
       number = {3},
        pages = {485-498},
          doi = {10.1093/pasj/58.3.485},
archivePrefix = {arXiv},
       eprint = {astro-ph/0512154},
 primaryClass = {astro-ph},
       adsurl = {https://ui.adsabs.harvard.edu/abs/2006PASJ...58..485T}
}

@ARTICLE{PyNeb,
       author = {{Luridiana}, V. and {Morisset}, C. and {Shaw}, R.~A.},
        title = "{PyNeb: a new tool for analyzing emission lines. I. Code description and validation of results}",
      journal = {\aap},
         year = 2015,
        month = jan,
       volume = {573},
          eid = {A42},
        pages = {A42},
          doi = {10.1051/0004-6361/201323152},
archivePrefix = {arXiv},
       eprint = {1410.6662},
 primaryClass = {astro-ph.IM},
       adsurl = {https://ui.adsabs.harvard.edu/abs/2015A&A...573A..42L}
}

@ARTICLE{Zier25,
       author = {{Zier}, Oliver and {Kannan}, Rahul and {Smith}, Aaron and {Puchwein}, Ewald and {Vogelsberger}, Mark and {Borrow}, Josh and {Garaldi}, Enrico and {Keating}, Laura and {McClymont}, William and {Shen}, Xuejian and {Hernquist}, Lars},
        title = "{The THESAN-ZOOM project: Population III star formation continues until the end of reionization}",
      journal = {\mnras},
         year = 2025,
        month = nov,
       volume = {544},
       number = {1},
        pages = {410-429},
          doi = {10.1093/mnras/staf1053},
archivePrefix = {arXiv},
       eprint = {2503.03806},
 primaryClass = {astro-ph.GA},
       adsurl = {https://ui.adsabs.harvard.edu/abs/2025MNRAS.544..410Z}
}

@ARTICLE{Yajima23,
       author = {{Yajima}, Hidenobu and {Abe}, Makito and {Fukushima}, Hajime and {Ono}, Yoshiaki and {Harikane}, Yuichi and {Ouchi}, Masami and {Hashimoto}, Takuya and {Khochfar}, Sadegh},
        title = "{FOREVER22: the first bright galaxies with Population III stars at redshifts z ≃ 10-20 and comparisons with JWST data}",
      journal = {\mnras},
         year = 2023,
        month = nov,
       volume = {525},
       number = {4},
        pages = {4832-4839},
          doi = {10.1093/mnras/stad2497},
archivePrefix = {arXiv},
       eprint = {2211.12970},
 primaryClass = {astro-ph.GA},
       adsurl = {https://ui.adsabs.harvard.edu/abs/2023MNRAS.525.4832Y}
}

@ARTICLE{Maiolino24,
       author = {{Maiolino}, Roberto and {{\"U}bler}, Hannah and {Perna}, Michele and {Scholtz}, Jan and {D'Eugenio}, Francesco and {Witten}, Callum and {Laporte}, Nicolas and {Witstok}, Joris and {Carniani}, Stefano and {Tacchella}, Sandro and {Baker}, William M. and {Arribas}, Santiago and {Nakajima}, Kimihiko and {Eisenstein}, Daniel J. and {Bunker}, Andrew J. and {Charlot}, St{\'e}phane and {Cresci}, Giovanni and {Curti}, Mirko and {Curtis-Lake}, Emma and {de Graaff}, Anna and {Egami}, Eiichi and {Ji}, Zhiyuan and {Johnson}, Benjamin D. and {Kumari}, Nimisha and {Looser}, Tobias J. and {Maseda}, Michael and {Nelson}, Erica and {Robertson}, Brant and {Rodr{\'\i}guez Del Pino}, Bruno and {Sandles}, Lester and {Simmonds}, Charlotte and {Smit}, Renske and {Sun}, Fengwu and {Venturi}, Giacomo and {Williams}, Christina C. and {Willmer}, Christopher N.~A.},
        title = "{JADES. Possible Population III signatures at z = 10.6 in the halo of GN-z11}",
      journal = {\aap},
         year = 2024,
        month = jul,
       volume = {687},
          eid = {A67},
        pages = {A67},
          doi = {10.1051/0004-6361/202347087},
archivePrefix = {arXiv},
       eprint = {2306.00953},
 primaryClass = {astro-ph.GA},
       adsurl = {https://ui.adsabs.harvard.edu/abs/2024A&A...687A..67M}
}

@ARTICLE{Vanzella23,
       author = {{Vanzella}, E. and {Loiacono}, F. and {Bergamini}, P. and {Me{\v{s}}tri{\'c}}, U. and {Castellano}, M. and {Rosati}, P. and {Meneghetti}, M. and {Grillo}, C. and {Calura}, F. and {Mignoli}, M. and {Brada{\v{c}}}, M. and {Adamo}, A. and {Rihtar{\v{s}}i{\v{c}}}, G. and {Dickinson}, M. and {Gronke}, M. and {Zanella}, A. and {Annibali}, F. and {Willott}, C. and {Messa}, M. and {Sani}, E. and {Acebron}, A. and {Bolamperti}, A. and {Comastri}, A. and {Gilli}, R. and {Caputi}, K.~I. and {Ricotti}, M. and {Gruppioni}, C. and {Ravindranath}, S. and {Mercurio}, A. and {Strait}, V. and {Martis}, N. and {Pascale}, R. and {Caminha}, G.~B. and {Annunziatella}, M. and {Nonino}, M.},
        title = "{An extremely metal-poor star complex in the reionization era: Approaching Population III stars with JWST}",
      journal = {\aap},
         year = 2023,
        month = oct,
       volume = {678},
          eid = {A173},
        pages = {A173},
          doi = {10.1051/0004-6361/202346981},
archivePrefix = {arXiv},
       eprint = {2305.14413},
 primaryClass = {astro-ph.GA},
       adsurl = {https://ui.adsabs.harvard.edu/abs/2023A&A...678A.173V}
}

@ARTICLE{Nakajima25,
       author = {{Nakajima}, Kimihiko and {Ouchi}, Masami and {Harikane}, Yuichi and {Vanzella}, Eros and {Ono}, Yoshiaki and {Isobe}, Yuki and {Nishigaki}, Moka and {Tsujimoto}, Takuji and {Nakamura}, Fumitaka and {Xu}, Yi and {Umeda}, Hiroya and {Zhang}, Yechi},
        title = "{An Ultra-Faint, Chemically Primitive Galaxy Forming at the Epoch of Reionization}",
      journal = {arXiv e-prints},
         year = 2025,
        month = jun,
          eid = {arXiv:2506.11846},
        pages = {arXiv:2506.11846},
          doi = {10.48550/arXiv.2506.11846},
archivePrefix = {arXiv},
       eprint = {2506.11846},
 primaryClass = {astro-ph.GA},
       adsurl = {https://ui.adsabs.harvard.edu/abs/2025arXiv250611846N}
}

@ARTICLE{Fujimoto25,
       author = {{Fujimoto}, Seiji and {Naidu}, Rohan P. and {Chisholm}, John and {Atek}, Hakim and {Endsley}, Ryan and {Kokorev}, Vasily and {Furtak}, Lukas J. and {Pan}, Richard and {Liu}, Boyuan and {Bromm}, Volker and {Venditti}, Alessandra and {Visbal}, Eli and {Sarmento}, Richard and {Weibel}, Andrea and {Oesch}, Pascal A. and {Brammer}, Gabriel and {Schaerer}, Daniel and {Adamo}, Angela and {Berg}, Danielle A. and {Bezanson}, Rachel and {Bouwens}, Rychard and {Chemerynska}, Iryna and {Claeyssens}, Ad{\'e}la{\"\i}de and {Dessauges-Zavadsky}, Miroslava and {Frebel}, Anna and {Korber}, Damien and {Labbe}, Ivo and {Marques-Chaves}, Rui and {Matthee}, Jorryt and {McQuinn}, Kristen B.~W. and {Mu{\~n}oz}, Julian B. and {Natarajan}, Priyamvada and {Saldana-Lopez}, Alberto and {Suess}, Katherine A. and {Volonteri}, Marta and {Zitrin}, Adi},
        title = "{GLIMPSE: An Ultrafaint ≃{}10$^{5}$ M$_{{\ensuremath{\odot}}}$ Pop III Galaxy Candidate and First Constraints on the Pop III UV Luminosity Function at z ≃ 6{\textendash}7}",
      journal = {\apj},
         year = 2025,
        month = aug,
       volume = {989},
       number = {1},
          eid = {46},
        pages = {46},
          doi = {10.3847/1538-4357/ade9a1},
archivePrefix = {arXiv},
       eprint = {2501.11678},
 primaryClass = {astro-ph.GA},
       adsurl = {https://ui.adsabs.harvard.edu/abs/2025ApJ...989...46F}
}

@ARTICLE{Cullen25,
       author = {{Cullen}, F. and {Carnall}, A.~C. and {Scholte}, D. and {McLeod}, D.~J. and {McLure}, R.~J. and {Arellano-C{\'o}rdova}, K.~Z. and {Stanton}, T.~M. and {Donnan}, C.~T. and {Dunlop}, J.~S. and {Shapley}, A.~E. and {Barrufet}, L. and {Begley}, R. and {Bondestam}, C. and {Cirasuolo}, M. and {Leung}, H.-H. and {Pollock}, C.~L. and {Stevenson}, S.},
        title = "{The JWST EXCELS survey: an extremely metal-poor galaxy at z = 8.271 hosting an unusual population of massive stars}",
      journal = {\mnras},
         year = 2025,
        month = jul,
       volume = {540},
       number = {3},
        pages = {2176-2194},
          doi = {10.1093/mnras/staf838},
archivePrefix = {arXiv},
       eprint = {2501.11099},
 primaryClass = {astro-ph.GA},
       adsurl = {https://ui.adsabs.harvard.edu/abs/2025MNRAS.540.2176C}
}

@ARTICLE{Pollock25,
       author = {{Pollock}, Clara L. and {Gottumukkala}, Rashmi and {Heintz}, Kasper E. and {Brammer}, Gabriel B. and {Roberts-Borsani}, Guido and {Oesch}, Pascal A. and {Witstok}, Joris and {Arellano-C{\'o}rdova}, Karla Z. and {Cullen}, Fergus and {Scholte}, Dirk and {Terp}, Chamilla and {Rowland}, Lucie and {Sneppen}, Albert and {Ito}, Kei and {Valentino}, Francesco and {Matthee}, Jorryt and {Watson}, Darach and {Toft}, Sune},
        title = "{Novel $z\sim~10$ auroral line measurements extend the gradual offset of the FMR deep into the first Gyr of cosmic time}",
      journal = {arXiv e-prints},
         year = 2025,
        month = jun,
          eid = {arXiv:2506.15779},
        pages = {arXiv:2506.15779},
          doi = {10.48550/arXiv.2506.15779},
archivePrefix = {arXiv},
       eprint = {2506.15779},
 primaryClass = {astro-ph.GA},
       adsurl = {https://ui.adsabs.harvard.edu/abs/2025arXiv250615779P}
}

@ARTICLE{Valentino23,
       author = {{Valentino}, Francesco and {Brammer}, Gabriel and {Gould}, Katriona M.~L. and {Kokorev}, Vasily and {Fujimoto}, Seiji and {Jespersen}, Christian Kragh and {Vijayan}, Aswin P. and {Weaver}, John R. and {Ito}, Kei and {Tanaka}, Masayuki and {Ilbert}, Olivier and {Magdis}, Georgios E. and {Whitaker}, Katherine E. and {Faisst}, Andreas L. and {Gallazzi}, Anna and {Gillman}, Steven and {Gim{\'e}nez-Arteaga}, Clara and {G{\'o}mez-Guijarro}, Carlos and {Kubo}, Mariko and {Heintz}, Kasper E. and {Hirschmann}, Michaela and {Oesch}, Pascal and {Onodera}, Masato and {Rizzo}, Francesca and {Lee}, Minju and {Strait}, Victoria and {Toft}, Sune},
        title = "{An Atlas of Color-selected Quiescent Galaxies at z > 3 in Public JWST Fields}",
      journal = {\apj},
         year = 2023,
        month = apr,
       volume = {947},
       number = {1},
          eid = {20},
        pages = {20},
          doi = {10.3847/1538-4357/acbefa},
archivePrefix = {arXiv},
       eprint = {2302.10936},
 primaryClass = {astro-ph.GA},
       adsurl = {https://ui.adsabs.harvard.edu/abs/2023ApJ...947...20V}
}

@MISC{Brammer_msaexp,
       author = {{Brammer}, Gabriel},
        title = "{msaexp: NIRSpec analyis tools}",
 howpublished = {Zenodo},
         year = 2023,
        month = sep,
          eid = {10.5281/zenodo.7299500},
          doi = {10.5281/zenodo.7299500},
      version = {0.6.17},
    publisher = {Zenodo},
       adsurl = {https://ui.adsabs.harvard.edu/abs/2022zndo...7299500B}
}

@ARTICLE{Miranda25,
       author = {{Miranda}, Henrique and {Pappalardo}, Ciro and {Afonso}, Jos{\'e} and {Papaderos}, Polychronis and {Lobo}, Catarina and {Paulino-Afonso}, Ana and {Carvajal}, Rodrigo and {Matute}, Israel and {Lagos}, Patricio and {Barbosa}, Davi},
        title = "{Importance of modelling the nebular continuum in galaxy spectra}",
      journal = {\aap},
         year = 2025,
        month = feb,
       volume = {694},
          eid = {A102},
        pages = {A102},
          doi = {10.1051/0004-6361/202451648},
archivePrefix = {arXiv},
       eprint = {2412.12060},
 primaryClass = {astro-ph.GA},
       adsurl = {https://ui.adsabs.harvard.edu/abs/2025A&A...694A.102M}
}

@ARTICLE{Cameron2024_NDG,
       author = {{Cameron}, Alex J. and {Katz}, Harley and {Witten}, Callum and {Saxena}, Aayush and {Laporte}, Nicolas and {Bunker}, Andrew J.},
        title = "{Nebular dominated galaxies: insights into the stellar initial mass function at high redshift}",
      journal = {\mnras},
         year = 2024,
        month = oct,
       volume = {534},
       number = {1},
        pages = {523-543},
          doi = {10.1093/mnras/stae1547},
archivePrefix = {arXiv},
       eprint = {2311.02051},
 primaryClass = {astro-ph.GA},
       adsurl = {https://ui.adsabs.harvard.edu/abs/2024MNRAS.534..523C}
}



\appendix

\section{Posterior results on the nebular fitting}
Here is provided the full corner plot of the nebula model described in Sect. \ref{ssec:neb}. 


\begin{figure}[!hb]
    \onecolumn
    \centering
    \includegraphics[width=0.8\linewidth]{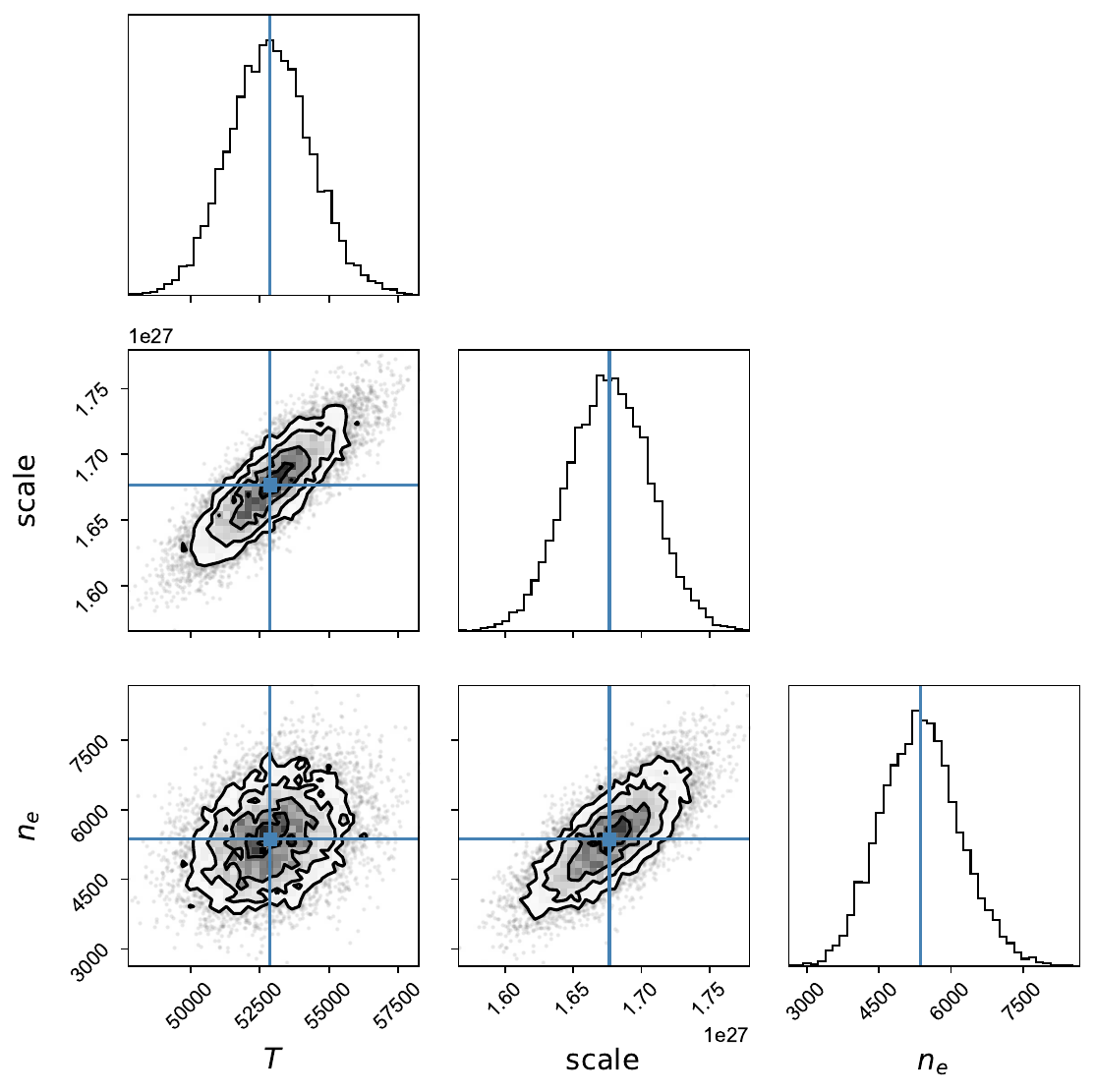}
    \caption{The posterior plot for the nebula model. All posterior distributions are clearly gaussian and the best fit parameters are well defined within the parameter space.}
    \label{fig:posterior}
\end{figure}

\end{document}